\documentclass{aa}  

\usepackage{graphicx}
\usepackage{xcolor}
\usepackage{soul}
\definecolor{v}{rgb}{0.54, 0.17, 0.89} 
\usepackage{txfonts}
\usepackage{hyperref}
\usepackage{orcidlink}

\definecolor{felipeA}{rgb}{0.75, 0.0, 0.2}
\usepackage{soul}
\usepackage{natbib}
\defcitealias{2022Wu}{Wu\&Shen22}
\defcitealias{2025Pan}{Pan+25}
\defcitealias{2025Bernal}{Bernal+25}
\defcitealias{2018SanchezSaez}{Sánchez-Sáez+18}
\begin{document}

   \title{Spectral Handling and Estimation of AGN Parameters \\ (SHEAP)}

   \subtitle{The first AGN fitting GPU-based code}

\author{F. Ávila-Vera\orcidlink{0000-0001-8332-7866}\inst{1}
\and
P. Sánchez-Sáez\orcidlink{0000-0003-0820-4692}\inst{2}
\and
V. Motta\orcidlink{0000-0003-4446-7465}\inst{1}
\and
S. Bernal\orcidlink{0000-0002-3519-3231}\inst{3,4}
}
   \institute{
Instituto de Física y Astronomía, Facultad de Ciencias, Universidad de Valparaíso,
Av. Gran Breta\~na 1111, Valparaíso, Chile\\
\email{felipe.avilav@postgrado.uv.cl, veronica.motta@uv.cl}
\and
European Southern Observatory, Karl-Schwarzschild-Str. 2, 85748 Garching, Germany
\and
Departamento de Astronomía, Universidad de Chile,
Camino El Observatorio 1515, Santiago, Chile
\and
Millennium Nucleus on Transversal Research and Technology to Explore Supermassive Black Holes (TITANS), Chile
}

   \date{Received --, --; accepted --,--}
 
 \abstract
{In the coming years, the number of discovered active galactic nuclei (AGN) is expected to significantly increase due to upcoming surveys. This rapid expansion of data will test our existing data analysis and modeling techniques and necessitate new methods for managing and interpreting large, complex datasets, particularly for next-generation surveys delivering heterogeneous spectra with a wide range of signal-to-noise ratios, spectral resolutions, and host-galaxy contamination.}
  {We present SHEAP (Spectral Handling and Estimation of AGN Parameters), a spectral fitting framework able to handle large quantities of data while being fast and reliable, without sacrificing physical interpretability, reproducibility, and robust uncertainty estimation for derived quantities.}
  {We use JAX, a Python GPU-powered framework, to implement a flexible modeling approach with modular components (continuum, host galaxy, Fe\,\textsc{ii} pseudo-continuum, and multi-component emission lines) and support for parameter tying and physically motivated constraints. We combine first-order gradient-based optimization with automatic differentiation to efficiently solve the highly complex problem, enabling stable convergence even in highly blended spectral regions (such as H\,$\beta$) and reducing runtime through vectorization and just-in-time compilation. }
{We compare our measurements against literature results and public fitting pipelines across four independent samples covering the C\,\textsc{iv}, Mg\,\textsc{ii}, H$\beta$, and H$\alpha$ spectral regions. We find generally good agreement for the main AGN spectral parameters, with typically $\sim 85$--$100\%$ of the objects lying within the $\pm0.3$ dex band, depending on the sample and parameter. The fits are statistically acceptable, with reduced chi-square distributions generally centered close to unity. Relative to the runtime reported by \citet{2026Bernal} using the \texttt{pPXF} framework, the fitting stage requires only $\sim1.7\%$ of the computational time, corresponding to an improvement of approximately $100$ times.}
{These results show that our method can deliver reliable AGN spectral decompositions with substantially lower computational cost than classical approaches. \texttt{SHEAP} is therefore well suited for the analysis of forthcoming massive spectroscopic datasets, where scalability, reproducibility, flexible model configurations, and robust uncertainty estimation are essential.}


\maketitle
%

\section{Introduction}
Since the discovery of active galactic nuclei (AGN) in 1960, they have opened a new field in astronomy \citep[for a historical review]{2015Kellermann}. These objects are characterized by strong emission across the entire electromagnetic spectrum, powered by the accretion of material onto a central supermassive black hole \citep[SMBH; see][for a detailed review]{2017Padovanib}.

Accretion not only powers the emission, but also gives rise to several substructures, including the broad-line region (BLR), a high-density, dust-free gas component moving at high velocity \citep[e.g., $\gtrsim 10^3~{\rm km\,s^{-1}}$][]{2011Shen,2017Rakshit}; the narrow-line region (NLR), a lower-velocity and lower-density ionized gas component \citep[e.g., $\lesssim 10^3~{\rm km\,s^{-1}}$][]{2013Mullaney,2017Calderone}; the dusty torus \citep{2013Assef}; and, in some systems, a central radio jet \citep[see][for a detailed discussion of these regions]{2015Netzer}. AGN are also classified into different types, largely interpreted within orientation-based unification schemes \citep{2017Padovani}. In this context, one of the most relevant distinctions is whether broad emission lines (i.e., evidence for the BLR) are present (type~I) or absent \citep[type~II;][]{1993Antonucci,1995Urry}.

Many of these components (e.g., the BLR and NLR, and accretion-disc emission) can be studied spectroscopically, as distinct spectral features trace different physical regions \citep{1992Boroson}. Nevertheless, this is not an easy task: optical spectra typically contain blended contributions from multiple components. For example, broad Fe\,\textsc{ii} multiplets can form a ``pseudo-continuum'' (the optical ``iron shelf'') that partially overlaps the Balmer lines and \ion{Mg}{ii}, complicating continuum placement and line deblending \citep[e.g.][]{1992Boroson,2004VeronCetty,2010Kovacevic}. In addition, non-virial kinematic components such as ionized outflows can produce asymmetric, blueshifted wings and extra broad components in narrow lines (e.g., \ion{O}{iii}$\lambda 5007$), biasing measurements if not modeled explicitly \citep[e.g.][]{2013Mullaney,2014Harrison,2016VillarMartin}. Finally, accretion-disc winds can imprint systematic blueshifts and broad absorption/emission signatures in high-ionization lines (classically in the UV, e.g., C\,\textsc{iv}), linking line profiles to large-scale mass loss and feedback \citep[e.g.][]{2003Crenshaw,2015King,2018Vietri}. Moreover, AGN optical spectra can be contaminated by host-galaxy light (e.g., stellar absorption features and a stellar continuum), which can dilute emission-line equivalent widths (EW) and bias continuum-based measurements if not properly accounted for \citep{2003Kauffmann}, adding another level of complexity to the physical interpretation of AGN spectra. These effects are particularly relevant for SMBH mass estimates based on virial broad-line methods, which rely on accurate measurements of the broad-line width and the AGN continuum luminosity \citep{2020OPopovic}.

Several tools have been developed over the years to identify AGN and to infer their physical properties, from the earliest spectroscopic classifications \citep{1974Khachikian} to modern large-scale characterization pipelines. Nevertheless, many analysis pipelines  have historically been developed for individual projects and have remained private \citep{2017Calderone}. This began to change over the last $\sim$15 years, as open-source practices became increasingly common in astronomy \citep{2009Weiner}, and multiple community tools have been released with different methodological approaches to model AGN spectra.

A milestone was the release of \texttt{QSFit} \citep{2017Calderone} and its Python implementation, \texttt{PyQSOFit} \citep{2018Guo}, which have been used to characterize large samples, including $750{,}414$ AGN spectra \citep{2022Wu} (hereafter, \citetalias{2022Wu}) observed with the SDSS \citep{2020Lyke}. Since then, additional tools have appeared to address specific challenges. For instance, because Fe\,\textsc{ii} emission can strongly affect continuum placement and measurements of blended lines, dedicated methodologies and automated pipelines such as \texttt{FANTASY} \citep{2020Ilic,2022Rakic,2023Ilic} have been developed. More broadly, the need for robust host--AGN separation and more realistic uncertainty estimates has motivated approaches that incorporate full stellar-continuum and kinematic modeling, such as Bayesian AGN Decomposition Analysis \citep[\texttt{BADASS};][]{2019Sexton,2021Sexton}. \texttt{BADASS} integrates Penalized PiXel-Fitting (\texttt{pPXF}) \citep{2004Cappellari,2012Cappellari,2017Cappellari,2023Cappellari}, a widely used tool for galaxy spectra, to model the host-galaxy component in AGN observations.  

The release of the Dark Energy Spectroscopic Instrument (DESI) is already delivering spectra at an unprecedented volume, with DR1 providing spectra for $\gtrsim 18$ million unique targets. This scale forces analysis frameworks to prioritize automation, robustness, and well-characterized systematics \citep[e.g., pipeline and value-added products,][]{2024Siudek}. It has also motivated survey-optimized approaches such as \texttt{FastSpecFit} \citep{2023Moustakas}, as well as AGN-focused decomposition efforts such as \texttt{DASpec} \citep{2024Du}, which incorporate instrument- and survey-level information directly into the fitting process. More broadly, next-generation multi-object facilities are expected to push these requirements even further \citep[e.g., the discussion of survey-scale spectroscopy and end-to-end data products in][]{2024Mainieri}.

Nevertheless, all of them have been used, and remain widely adopted, in a variety of scientific applications \citep{2026Bernal,2026Hutsemekers,2025Layek,2024Ren,2023Fu}, each of them presents potential drawbacks: some rely on specific minimizers that may not be optimal for uncommon object populations; others lack flexibility in the choice of components (e.g., templates or parametric components); some have been developed primarily for a particular instrument or type of dataset; and others lack native value-added products\footnote{e.g., line flux, EW, luminosities, and SMBH mass}, as well as the ability to efficiently process enormous quantities of data (i.e., scalability). 

In an era in which next-generation surveys demand not only fast methodologies, but also flexible and physically motivated modeling frameworks to interpret heterogeneous AGN populations, these limitations may become significant. This becomes even more relevant with the start of 4MOST operations, which will further extend the observational landscape by delivering an exceptionally large AGN census: the 4MOST AGN survey is designed to provide physical properties for up to $\sim 10^{6}$ AGN, enabling population-scale tests of SMBH growth and AGN--galaxy co-evolution \citep{2019deJong,2019Merloni}. Within this context, the Chilean AGN/Galaxy Extragalactic Survey (ChAnGES) will provide complementary, targeted science cases \citep{2023Bauer}, focusing on uncommon AGN populations.

In parallel, high-performance computing, including GPU acceleration, has become increasingly important for large-scale inference and model evaluation in astronomy \citep[e.g.,][]{2022Galan,2022Gu,2023Michalewicz,2025Wang}, in line with broader ``astroinformatics'' and big-data trends in survey science \citep[e.g.,][]{2009Borne,2015Zhang,2020Vavilova}. Such developments provide a practical route to meet the computational demands posed by the rapidly increasing volume of spectroscopic data expected in the coming years.

In this context, we present \texttt{Spectral Handling and Estimation of AGN Parameters} (\texttt{SHEAP})\footnote{\url{https://sheap.readthedocs.io/}}, a GPU-enabled spectral decomposition tool designed to analyse large samples of optical AGN spectra, as well as spectra of inactive galaxies, and to deliver reliable parameter estimates with substantially reduced runtimes. To test \texttt{SHEAP}, we apply it to $\sim 2000$ AGN spectra, performing a consistent multi-component decomposition over a rest frame broad wavelength range ($1100$--$7000$~\AA) and across different AGN populations and spectral resolutions.

The paper is structured as follows. In Section~\ref{sec:Methods}, we describe our methodology. In Section~\ref{sec:test_code}, we present the validation tests used to assess the reliability of the code. In Section~\ref{sec:Discussion}, we discuss our results and conclusions.  Through this work we adopt a flat $\Lambda$CDM cosmology with $\Omega_\Lambda = 0.7$, $\Omega_\mathrm{M} = 0.3$, and $H_0 = 70~\mathrm{km\,s^{-1}\,Mpc^{-1}}$.

\section{Methods}
\label{sec:Methods}
In this section, we present and outline the different steps and strategies in our approach to modeling AGNs spectra using \texttt{SHEAP}, a Python 3 package powered by \texttt{JAX}~\citep{2021Bradbury}. 
\subsection{Data pre-processing}
Before delving into the modeling procedure of our method, it is important to explore the data structure and the corrections required to obtain physically representative results.
\subsubsection{Data handling \& structure }
\label{sec:DataSheap}
Our method operates on data stored as a three-dimensional array
$A_{i,j,k}$ , where:
\begin{itemize}
  \item $i \in \{0, \ldots, N-1\}$ indexes the spectra in the batch,
  \item $j \in \{0, \ldots, 2\}$ indexes the data component, corresponding to
        wavelength, flux density, and per-pixel observational uncertainty $(\sigma)$,
  \item $k \in \{0, \ldots, L-1\}$ indexes pixels within a spectrum.
\end{itemize}
Given the complexity of spectral modeling \citep[e.g.][]{2017Cappellari,2018Guo,2023Ilic}, an array-based, batched data layout enables efficient batch processing and GPU-based vectorization during model evaluation and optimization. By expressing the computation in terms of array-shaped operations, the data-parallel structure of the problem is made explicit to the underlying compiler and accelerator backends. This facilitates optimized memory access patterns \cite[for more details, see][]{2021Bradbury} and significantly accelerates both forward-model evaluation and automatic differentiation, particularly in large parameter spaces and in workflows that require repeated evaluations, such as gradient-based optimization and posterior sampling \citep[as in our case; see Section~\ref{sec:optstragety}][]{2017Dalalyan,2019Ma}.

\subsubsection{Spectral corrections}
We first apply the standard cosmological corrections to our sample\footnote{$\lambda_{\rm rest} = \lambda_{\rm obs}/(1+z)$ and $F_{\lambda}^{\rm rest} = (1+z)F_{\lambda}^{\rm obs}(\lambda_{\rm obs})$} \citep{2002Hogg,2007Blanton}. Since the observed spectra are also affected by Galactic dust extinction \citep{1999Fitzpatrick}, given the object coordinate we correct for extinction using E$(B{-}V)$ values, computed with the \texttt{sfdmap}\footnote{https://github.com/kbarbary/sfdmap} Python package, based on the full-sky dust maps of \citet{1998Schlegel}. The extinction correction is then applied using the reddening law of \citet{1989Cardelli} with $R_V = 3.1$, together with the recalibration of \citet{2011Schlafly}.


\subsection{Modeling with \texttt{SHEAP}}
\label{sec:modeling_in_sheap}
The \texttt{SHEAP} modeling pipeline constructs a spectral model (SM) that depends on a defined wavelength range, this interval determines the set of physical components that will appear in an AGN spectrum model as host-galaxy starlight, Fe\,\textsc{ii} emission, non-thermal continuum, and emission lines (EL). Then, for a general spectral model:
\begin{align}
\label{eq:spectralmodel}
F_{\mathrm{SM}}(\lambda_{\mathrm{rest}})
&=
F_{\mathrm{host}}(\lambda_{\mathrm{rest}})
+
F_{\mathrm{Fe}}(\lambda_{\mathrm{rest}}) \nonumber \\[4pt]
&\quad+
F_{\mathrm{cont}}(\lambda_{\mathrm{rest}})
+
F_{\mathrm{Balmer}}(\lambda_{\mathrm{rest}}) 
+
F_{\mathrm{EL}}(\lambda_{\mathrm{rest}}),
\end{align}
\noindent
where $F_{\mathrm{SM}}(\lambda_{\mathrm{rest}})$ is the total modeled flux density of the spectrum in the rest frame. The individual terms represent the different spectral components contributing to the observed emission: $F_{\mathrm{host}}(\lambda_{\mathrm{rest}})$ is the stellar continuum from the
host galaxy, $F_{\mathrm{Fe}}(\lambda_{\mathrm{rest}})$ accounts for the blended Fe\,\textsc{ii} emission complexes, $F_{\mathrm{cont}}(\lambda_{\mathrm{rest}})$ represents the featureless AGN continuum emission, $F_{\mathrm{Balmer}}(\lambda_{\mathrm{rest}})$ describes the Balmer continuum emission, and $F_{\mathrm{EL}}(\lambda_{\mathrm{rest}})$ represents the sum of all emission-line component (e.g narrow, broad, outflow, Winds).

\subsubsection{The host galaxy model}
\label{sec:host}
It is well known that, at $z \lesssim 1$, the stellar continuum and absorption features of the host galaxy can contribute significantly to the observed AGN spectrum \citep[e.g.,][]{2005Greene,2023Jalan,2024Ren}. To account for this effect, we implemented a lightweight host-galaxy module inspired by the penalized Pixel-Fitting (\texttt{pPXF}) methodology \citep{2004Cappellari,2017Cappellari}, designed specifically for AGN+host spectra.

In our approach, the host-galaxy stellar emission is modeled as a linear combination of single stellar population (SSP) templates spanning a grid of stellar ages and metallicities \citep{2010Vazdekis}. The composite template is convolved with a Gaussian line-of-sight velocity distribution (LOSVD) to reproduce the broadening induced by the stellar motions within the galaxy \citep{1993vanderMarel}. The convolution is performed efficiently in Fourier space \citep{1974Brigham}, ensuring consistent spectral broadening while keeping the kinematic description minimal. 

In contrast to full spectral-fitting frameworks such as \texttt{pPXF}, which can simultaneously recover detailed stellar kinematics and population properties using penalized likelihoods and Gauss--Hermite expansions \citep[e.g.,][]{2004Cappellari,2005CidFernandes}, our current implementation restricts the LOSVD to a purely Gaussian kernel. In this approximation, the LOSVD is parameterized only by the mean line-of-sight velocity $V$ and the velocity dispersion $\sigma$, without including higher-order Gauss--Hermite moments. Similarly, the method does not attempt to infer non-parametric star-formation histories. This design choice prioritizes numerical stability and computational speed in regimes where the host contribution is typically subdominant and strongly blended with the AGN continuum and emission-line components. 

We built a custom cube of SSP templates from E-MILES \citep{2016Vazdekis} using the ``Tune SSP Models'' web tool\footnote{\url{https://research.iac.es/proyecto/miles/pages/webtools/tune-ssp-models.php}}. The original library spans $1{,}680$--$49{,}999.4$\,\AA; which we selected the $1{,}680$--$8{,}950$\,\AA\ interval, representative of the typical wavelength coverage of UV/optical instruments. For consistency, we adopted the values used in \citet{2023Cappellari}. Accordingly, the templates were generated with linear sampling, $\Delta\lambda = 0.9$\,\AA\,pix$^{-1}$, and have a native spectral resolution of $\mathrm{FWHM}\simeq 2.51$--$5$\,\AA\ across the selected range. We used the Padova+00 \citep{2000Girardi} isochrones, a unimodal IMF with slope $2.30$, the base Fe abundance pattern, and the full E-MILES metallicity ([M/H] $= -2.32$ to $0.22$) and age ($0.1$--$10$\,Gyr) grids.

\subsubsection{The AGN Continuum model}

Our method provides flexibility in the continuum component, allowing the background model to be chosen according to the physical nature of the source and the wavelength range under analysis. Among the available options are a linear baseline for small spectral windows with nearly flat local continua \citep{2001VandenBerk}, a power-law continuum for the featureless accretion-disk emission commonly observed in AGN \citep[e.g.,][]{2006Richards}, and a broken power law for wider wavelength ranges where a single slope does not adequately describe the continuum shape \citep{2008Dong}.


\subsubsection{The Balmer emission model}
\label{sec:balmeremission}
We model the Balmer emission as an independent set components \citep[][]{2000Sulentic} that shape the pseudo-continuum in the near-ultraviolet (UV) region.
The Balmer continuum, produced by bound--free recombination of hydrogen in partially optically thick gas, becomes significant shortward of the Balmer edge at 3,646\,\AA\ \citep{2000Sulentic,2011Dong}.

Meanwhile, at longer wavelengths (3,700--4,000\,\AA) emission from high-order Balmer lines
(typically $n \gtrsim 7$) blends together due to Doppler broadening and instrumental resolution, forming a quasi-continuum that further elevates the background level.
This blended Balmer emission is particularly prominent in luminous and high-redshift AGNs, where it can substantially affect the apparent continuum shape and the measurement of nearby spectral line features \citep{2002Dietrich}. 

Formally, we split the total Balmer contribution as:
\begin{equation}
F_{\rm Balmer}(\lambda_{\rm rest}) \;=\;
F_{\rm PB\,}(\lambda_{\rm rest})
\;+\;
F_{\rm HighBalmer}(\lambda_{\rm rest}),
\end{equation}

\noindent
where $F_{\rm PB}(\lambda_{\rm rest})$ represents a parametric model of the Balmer continuum, defined as in \cite{1982Grandi,2002Dietrich}, and $F_{\rm HighBalmer}(\lambda_{\rm rest})$ accounts for the high-order Balmer pseudo-continuum.

On the other hand, the high-order Balmer component is modeled using an empirical template, $T_{\rm HBalmer}(\lambda)$, constructed from the blend of high-$n$ Balmer emission lines that merge blueward of 4\,000\,\AA\ \citep{1995Storey}. In this work, we adopt the updated version of this template presented in \cite{2025Bernal} (hereafter, \citetalias{2025Bernal}), which extends the modeled series to $n > 7$. The template is first convolved with a Gaussian kernel to reproduce the observed line broadening, and it may subsequently be shifted in velocity and scaled by a multiplicative factor to match the observed spectrum.


\subsubsection{The Fe\,\textsc{ii} treatment}

The Fe\,\textsc{ii} emission is a prominent source of contamination and degeneracy in the study of AGN spectra, arising from a multitude of blended transitions of singly ionized iron. It spans a broad wavelength range, from the ultraviolet \citep[e.g., around Mg\,\textsc{ii};][]{2001Vestergaard,2019Popovic}, through the optical \citep[e.g., near H$\beta$ and H$\alpha$;][]{2004VeronCetty,2022Park}, to the near-infrared \citep[e.g., around Pa$\gamma$;][]{2000Rudy,2008Landt,2012GarciaRissmann}. This pseudo-continuum frequently overlaps with key diagnostic lines and can significantly bias measurements of broad-line widths and fluxes if not properly accounted for. Accurate modeling and subtraction of both Fe\,\textsc{ii} and Fe\,\textsc{iii} emission are therefore essential for recovering the intrinsic properties of the BLR and for robustly measuring the Balmer and Mg\,\textsc{ii} line parameters.

To model this component, we adopt the classical template-fitting approach, in which an empirical Fe template is scaled, broadened through convolution with a Gaussian kernel, and allowed to shift in velocity to match the data, following the same procedure described in Section~\ref{sec:balmeremission}. In particular, we use the Fe,\textsc{ii} templates from \cite{1992Boroson} and \cite{2001Vestergaard}, following implementations such as \citet{2017Calderone}, which provide continuous coverage from 1,074 to 7,000~\AA\ by combining UV and optical Fe emission complexes. In this scheme, no explicit separation into broad'' and narrow'' Fe components is introduced; instead, the template is treated as a single blended complex whose effective kinematics are captured by the broadening and velocity shift. Nevertheless, we note that the use of templates may be insufficiently flexible in particularly complex regions, such as the vicinity of H$\alpha$. For this reason, we also implemented the empirical Fe modeling strategy available in \texttt{FANTASY}, which includes a full model of Fe\,\textsc{ii} emission based solely on atomic data. This model includes a total of 283 transitions, divided into 17 atomic groups, over the wavelength range 3,700--11,000~\AA\ \citep[see][for details]{2023Ilic}. In this approach, pseudo-templates are generated for the different Fe atomic groups and tied together in velocity shift and width.

\subsubsection{Emission-lines treatment}
\label{sec:EmissionLinesModels}
Emission-line fitting is a central component of AGN spectral modeling, since line profiles trace the kinematic signatures of different emitting regions, including broad and narrow components, as well as winds and outflows. Our method is designed to efficiently fit complex models comprising multiple profile families (e.g., Gaussian, Lorentzian, and Voigt), multiple kinematic components within a given line, and multiple spectral regions simultaneously. It also accounts for the kinematics of Fe\,\textsc{ii} emission, the stellar component, and Balmer emission. As part of the modeling strategy, we adopt a predefined list of emission lines that distinguishes between transitions known to exhibit broad emission, transitions typically associated only with narrow emission, and transitions that can display both narrow and broad components \citep{2023Ilic}. This scheme helps define a consistent and physically motivated line configuration for each fitted region. The parameter bounds for these regions are set according to the values presented in Table~\ref{tab:region_limits}, ensuring that the fitted parameters remain physically meaningful. In addition, we incorporate a set of known atomic constraints into the main fitting structure, as in the cases of \mbox{[O\,\textsc{iii}]} and \mbox{[N\,\textsc{ii}]} \citep{2006Vestergaard,2016MejiaRestrepo}.

\subsection{Optimization Strategy}
\label{sec:optstragety}
Over the past decades, optimization strategies for non-linear spectral fitting have evolved substantially. Early and widely adopted approaches rely on classical non-linear least-squares solvers such as the Levenberg--Marquardt (LM) algorithm \citep{1978More}, which remains a standard choice in many fitting packages (e.g., \texttt{LMFIT}; \citealt{2014Newville}) and \texttt{MPFIT} \citep{2009Markwardt}. In parallel, posterior exploration via Monte Carlo Markov Chain (MCMC) samplers, such as \texttt{emcee} \citep{2013ForemanMackeycode,2013ForemanMackey}, has become common when robust uncertainty quantification is required. More recently, constrained non-linear least-squares minimizers have also been introduced in widely used spectral-fitting frameworks (e.g., \texttt{capfit} within \texttt{pPXF}; \citealt{2023Cappellari} and references therein), reflecting the need to handle increasingly flexible and high-dimensional models.

In this work, we adopt a first-order, gradient-based optimization strategy within the \texttt{JAX} framework. With the model defined for the different spectral regions (Eq.~\ref{eq:spectralmodel}), we impose physically motivated bounds that depend on the kinematic component, summarized in Table~\ref{tab:region_limits}. The optimization in \texttt{SHEAP} is performed using the Adam optimizer \citep{2014Kingma}, implemented through the \texttt{Optax} library \citep{deepmind2020jax}, that depends on two hyper parameters: the learning rate and the number of iterations. This choice enables efficient and adaptive gradient updates, while leveraging automatic differentiation and accelerated compilation for fast model evaluation and differentiation.

Our minimization strategy accounts for fixed and bounded parameters (as well as tied relations where applicable). Parameter bounds are enforced via smooth re-parameterizations to an unconstrained optimization space, avoiding hard projections that can hinder gradient-based methods. This approach is conceptually similar to the bound-handling strategy adopted in \texttt{LMFIT} \citep{2014Newville}, where bounds are implemented through internal parameter transformations. This design keeps the optimizer agnostic to the underlying spectral model and ensures that constraints are applied consistently across components and regions.

Residuals are evaluated in a weighted manner using the spectral uncertainties, so that high-S/N pixels carry greater influence in the fit. The normalized residual vector is defined as

\begin{equation}
r_i = \frac{f_{\mathrm{model},i} - f_{\mathrm{obs},i}}{\sigma_i},
\end{equation}
where $f_{\mathrm{model},i}$ corresponds to the model evaluated at pixel $i$, $ f_{\mathrm{obs},i}$ is the observed flux at pixel $i,$ and $\sigma_i$ is the uncertainty by pixel. We adopt a robust residual loss based on the log-cosh function,
\begin{equation}
\mathcal{L}_{\mathrm{residual}} =
\left\langle \log\cosh\left(r_i\right) \right\rangle .
\end{equation}
The log-cosh function provides a smooth and outlier-tolerant alternative to a purely quadratic ($L_2$) loss: it behaves approximately quadratically for small residuals, while transitioning to an approximately linear ($L_1$-like) penalty at large $|r_i|$. As a result, isolated deviant pixels (e.g., residual sky features, imperfect masking, or artifacts) have reduced leverage on the solution, while the loss remains differentiable and therefore well-suited for gradient-based optimization \citep[see][for a detailed discussion]{2022Saleh}.

It is important to highlight that this procedure is applied consistently to the full batch of fitted spectra, using the same model configuration for all objects. Because \texttt{SHEAP} is implemented to run natively in parallel (i.e., the same computation is evaluated over a batch of spectra in a single compiled call), the optimization is carried out simultaneously across the batch rather than spectrum-by-spectrum.

\begin{figure*}[h]
    \centering
    \includegraphics[width=.75\linewidth]{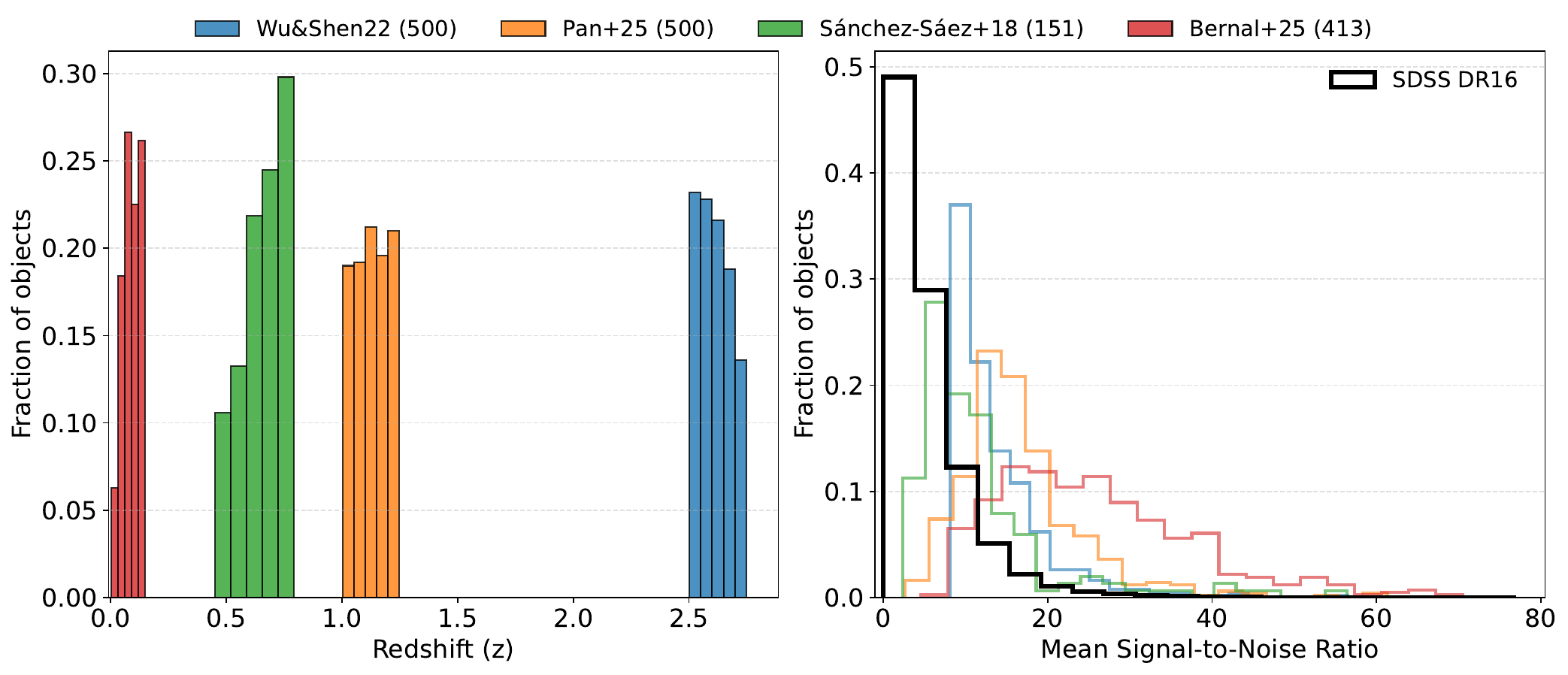}
    \caption{Left: Redshift distribution by sample. Right: Mean signal-to-noise ratio (S/N) distribution by sample. Sample names are shown at the top of the figure, with the number of objects in parentheses, and the distribution of  S/N in DR16 has an example of the expected distribution}
    \label{fig:redshift_snr_fraction}
\end{figure*}
\subsection{Uncertainty Estimation}
\label{sec:uncertainty}

Because AGN spectral decomposition is intrinsically nonlinear, the best-fit parameters alone may not capture the full correlation structure among parameters, nor the impact of local degeneracies between continuum components and emission-line profiles. To estimate uncertainties beyond local curvature approximations, \texttt{SHEAP} provides a Monte Carlo (MC) resampling mode based on perturbation sampling \citep[e.g.,][]{1992Press}. In this approach, the observed spectrum is repeatedly perturbed according to its per-pixel uncertainties, and the full model is re-fit for each perturbed realization. The ensemble of re-fits yields an empirical distribution of model parameters that naturally captures parameter covariances and certain nonlinear effects \citep{2010Andrae,2011Shen}.

Each MC realization is initialized at the same initial conditions as the main fitting. This choice accelerates convergence while still allowing the optimizer to explore nearby basins of the loss landscape when the likelihood surface is multimodal or exhibits strong degeneracies. Similar ``flux-randomization + refitting'' strategies are commonly used in quasar spectral fitting pipelines to obtain robust measurement uncertainties \citep[e.g.,][]{2018Guo,2022Wu}.

\subsection{Post-fit derived quantities}
\label{sec:postfitting}

Beyond delivering a single best-fitting model, \texttt{SHEAP} is designed to support a consistent post-processing stage in which both best-fit parameters and sampled parameters can be propagated into physically relevant (``added-value'') quantities. In practice, the code can (i) evaluate derived quantities directly from the best-fit solution and/or (ii) evaluate them for each sample obtained (see Section~\ref{sec:uncertainty}), thereby producing posterior distributions for quantities such as line fluxes and luminosities, monochromatic continuum luminosities, EW, and measurements of post-processed (combined) broad-line profiles.

This design ensures that uncertainties in derived quantities are derived from the same modeling assumptions used during the fit (e.g., the adopted continuum and pseudo-continuum definitions, kinematic ties, and line-profile parameterization), while enabling downstream analyses to use either point estimates or full posterior samples. In the next sections, we explain how these are defined.

\subsubsection{ Additional products derived by \texttt{SHEAP}}
\label{sec:extraproducts}

A central goal of \texttt{SHEAP} is to translate spectral fits into physically meaningful and reproducible measurements. For each fitted spectrum and, when available, for each MC realization, the code provides a set of value-added products, including:

\begin{itemize}
\item Emission-line products (per line and, when applicable, for post-processed combined profiles):
integrated line fluxes and line luminosities; EW; and kinematic/profile measurements such as FWHM and velocity shift, together with the other line parameters returned by the fit (e.g., amplitudes and profile-shape parameters, depending on the adopted parameterization).

\item Monochromatic continuum luminosities: $\lambda L_{\lambda}$ evaluated at the standard wavelengths used in AGN applications (e.g., 5{,}100, 3{,}000, and 1{,}350\,\AA) 
\end{itemize}

For emission lines, EW is computed using the standard definition relative to a local continuum level. In a decomposition context, this pseudo-continuum must be defined consistently (e.g., an AGN power-law plus additional continuum-like components such as Fe, host-galaxy starlight, and/or the Balmer continuum, depending on the adopted model) \citep[e.g.,][]{2017Calderone,2025Bernal}. For cosmological distances, we use the cosmological module of \texttt{astropy} \citep{2013Astropy,2018Astropy,2022Astropy}.

Our method also allows the main broad-line component to be estimated using different approaches, either by combining all broad components or by combining only those that are kinematically consistent (see Appendix~\ref{sec:broadcombination}).

In this way, our uncertainty-estimation framework (see Section~\ref{sec:uncertainty}) provides a natural mechanism for propagating uncertainties to all post-processing quantities, including luminosities, EW, combined-profile FWHM, and single-epoch SMBH mass estimates.
 
\begin{figure*}[h]
    \centering
    \includegraphics[width=.75\linewidth]{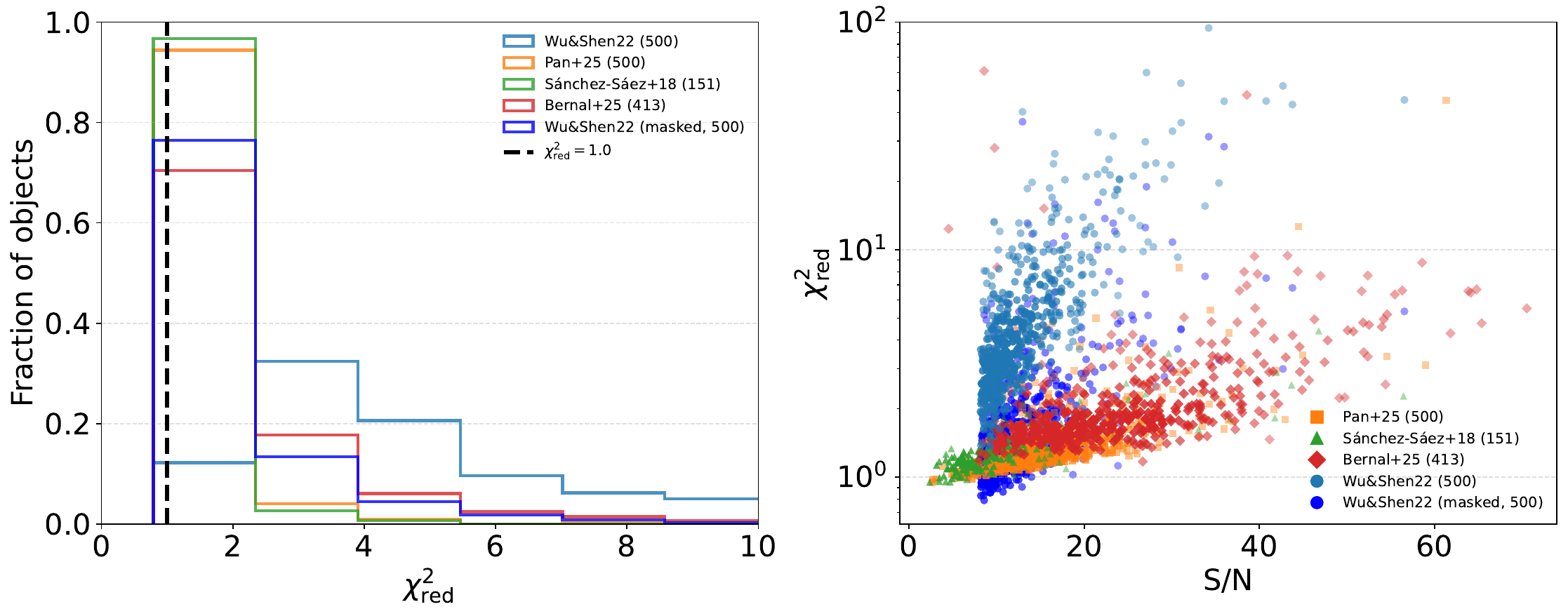}
    \caption{Right: Distribution of reduced $\chi^2$ by sample, normalized to the fraction of objects in each sample. The vertical step line marks $\chi^2_{\rm red}=1$. Left:
Relation between signal-to-noise ratio (S/N) and the reduced chi-square, $\chi^2_{\rm red}$. Each color represent a different sample.
}
    \label{fig:SNchi2plots}
\end{figure*}

\section{Performance validation of \texttt{SHEAP}}  
\label{sec:test_code}

To evaluate the performance of our method across different spectral regions, resolutions, and AGN populations, we perform a series of tests by comparing our results with those from previously published studies. To avoid bias toward any single methodology, we include comparisons spanning a variety of independent approaches. 

In these tests, all emission-line profiles are modeled with Gaussian components, and the full list of emission lines used in each model is given in Table~\ref{tab:emissionline}. The underlying AGN continuum is represented by a power law and, as in \citet{2023Ilic}, we adopt a single baseline model for each sample. This ensures that differences within each sample primarily reflect data quality and population diversity, rather than changes in the model configuration.

For the results presented in this paper, we optimize each model for 2000 iterations using a learning rate of 0.01 (see Section~\ref{sec:optstragety}). Uncertainties are estimated using our framework for 50 realizations (see Section~\ref{sec:uncertainty}).

For each compared parameter ($X$), we define the logarithmic difference as
\begin{equation}
    \Delta_X = \log_{10}(X_{\rm SHEAP}) - \log_{10}(X_{\rm lit}),
\end{equation}
and quantify the agreement using the fraction of objects satisfying $|\Delta_X|\leq0.3$ dex. We complement this metric with the median offset and the normalized median absolute deviation (NMAD), which characterize the systematic and random differences between our measurements and those from the literature.

All computations were performed on a workstation equipped with two Intel Xeon Gold 5220R (2.20,GHz) CPUs and an NVIDIA RTX A5000 GPU.

In Fig.~\ref{fig:redshift_snr_fraction}, we present the signal-to-noise ratio distributions for the different samples, together with their corresponding redshift distributions.

\subsection{The Wu\&Shen22 sample}  
\label{sec:WuandShen}
We first test the performance of \texttt{SHEAP} against \texttt{PyQSOFit} \citep{2018Guo}, as applied by \citetalias{2022Wu} to the DR16Q catalog of $750{,}414$ broad-line quasars over $0.1 \le z \le 6$. We select a stratified subsample with $2.0 \le z \le 2.25$ to test the C\,\textsc{iv} region, corresponding to a rest-frame wavelength coverage of 950--2{,}968~\AA. To reduce redshift-related systematics, we require consistency among the catalog redshift estimates, enforcing $|\Delta z| \le 0.01$.

As described in Section~\ref{sec:DataSheap}, \texttt{SHEAP} operates on array-based inputs. Since SDSS spectra can have different wavelength coverages and array lengths \citep{2012Bolton}, we resample the selected spectra onto a common wavelength grid using \texttt{SpectRes} \citep{2017Carnall}, which conserves flux density and propagates uncertainties during rebinning. For this test, we fit the 1,100--2,750~\AA\ range using a power-law continuum and Gaussian emission-line components, adopting three broad components for the C\,\textsc{iv} region, resulting in a model with 41 parameters.

Examples of the best-fit models are shown in Figs.~\ref{fig:spectra_wushen_median_snr} and \ref{fig:spectra_wushen_median_chi2}. The main source of increased $\chi^2_{\rm red}$ is absorption from the Ly$\alpha$ forest at $\lambda_{\rm rest}\lesssim1300$~\AA. After masking this region, the $\chi^2_{\rm red}$ distribution becomes significantly narrower, with the median and scatter decreasing from $4.3 \pm 3.7$ to $1.5 \pm 0.95$ (Fig.~\ref{fig:SNchi2plots}). We find no strong dependence of fit quality on S/N, although the highest-S/N objects show slightly larger $\chi^2_{\rm red}$ values.

\begin{figure}
    \centering
    \includegraphics[width=0.75\linewidth]{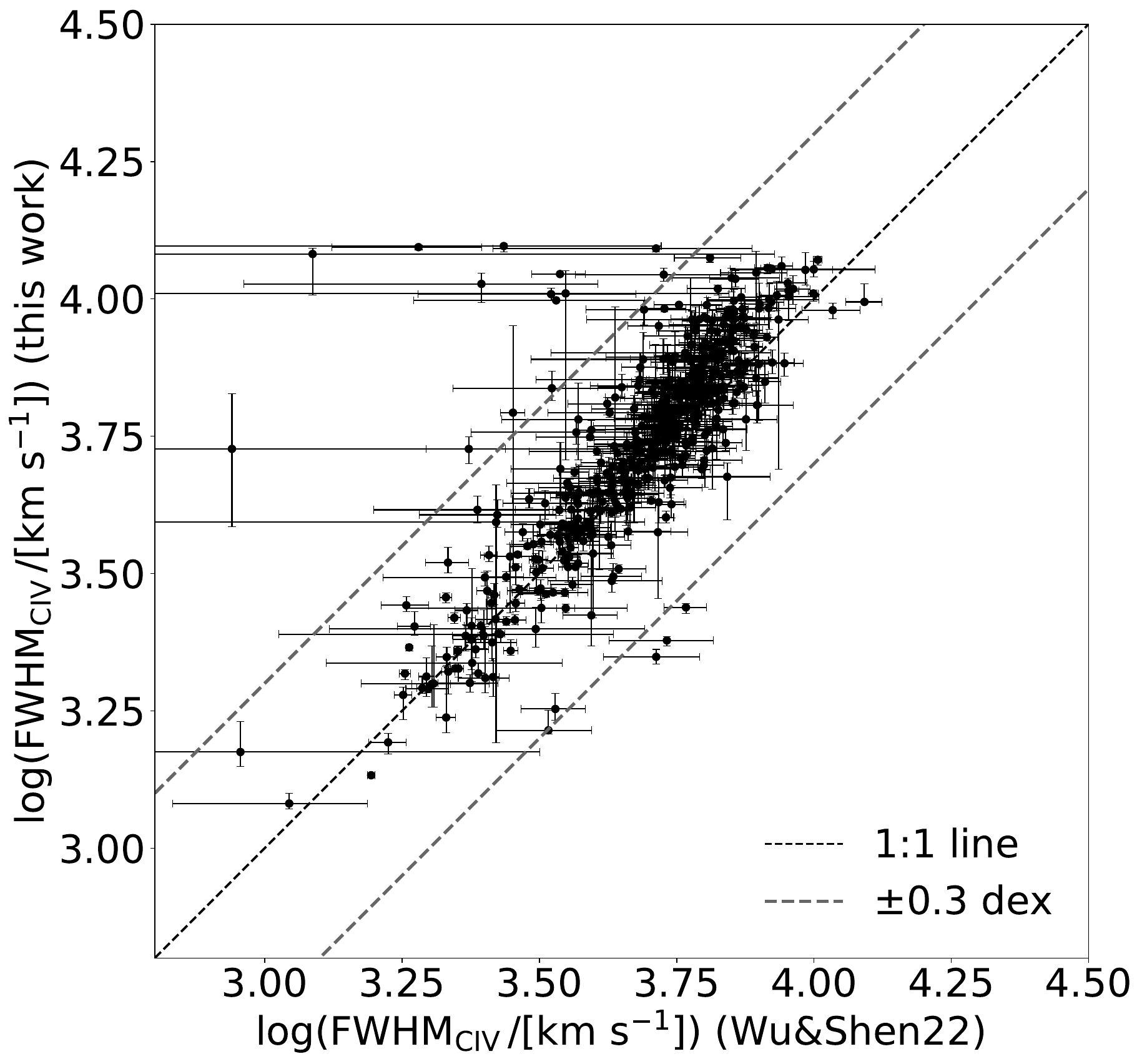}
    \vspace{0.25cm}
    \includegraphics[width=0.75\linewidth]{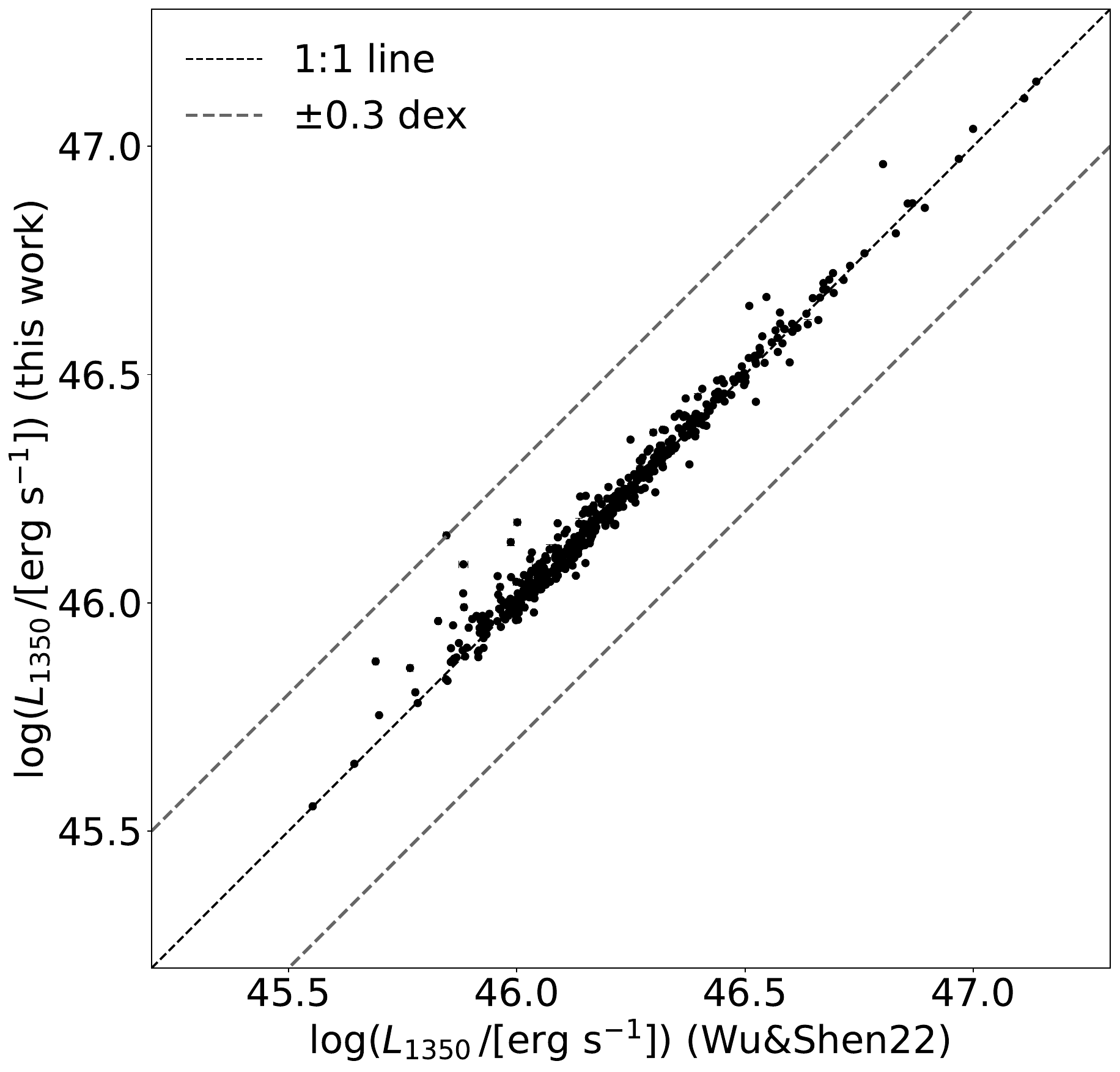}

    \caption{Comparison between our measurements and those of \citetalias{2022Wu}. Top: C\,\textsc{iv} emission-line FWHM in logarithmic scale. Bottom: Monochromatic continuum luminosity at 1350\,\AA\ in logarithmic scale. In both panels, the black dashed line indicates the 1:1 relation, while the gray dashed lines mark the $\pm 0.3$ dex region.}
    \label{fig:2022WuComparison}
\end{figure}

We compare our measurements with the C\,\textsc{iv} FWHM and $L_{1350}$ values reported by \citetalias{2022Wu}, combining broad components following the same strategy described in Appendix~\ref{sec:combiningall}. The results are shown in Fig.~\ref{fig:2022WuComparison}. For $\mathrm{FWHM}_{\mathrm{C\,IV}}$, 96.40\% of the objects lie within the $\pm0.3$ dex band, with a median bias of $+0.04$ dex and $\mathrm{NMAD}=0.07$ dex. For $L_{1350}$, the agreement is tighter, with 99.8\% of objects within $\pm0.3$ dex, a median bias of $+0.01$ dex, and $\mathrm{NMAD}=0.02$ dex. The broader scatter in $\mathrm{FWHM}_{\mathrm{C\,IV}}$ is expected because the broad profile is modeled with multiple components that are later combined.

\begin{figure}
    \centering
    \includegraphics[width=.85\linewidth]{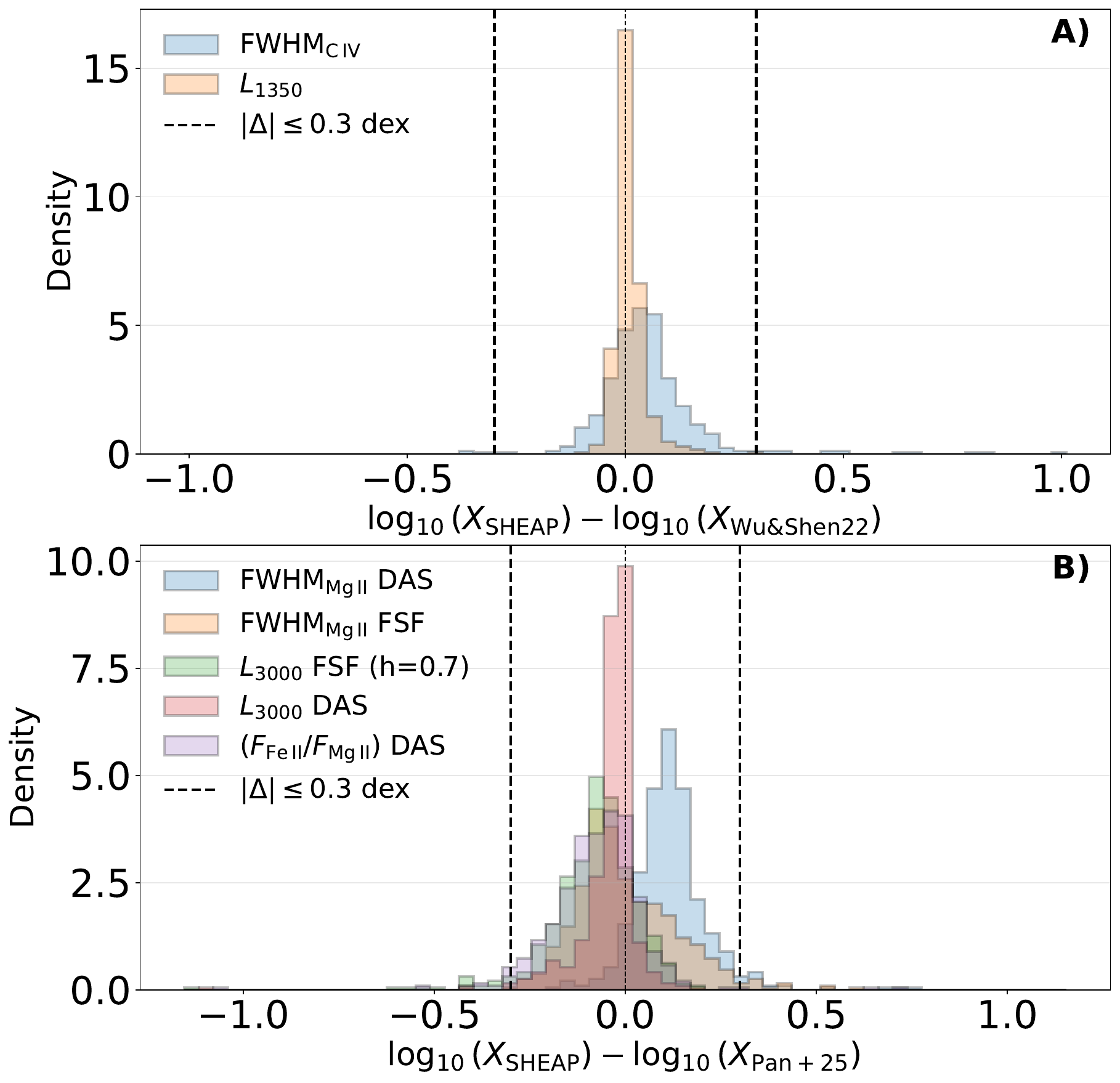}
  \caption{
Distributions of the logarithmic differences, $\log_{10}(X_{\rm SHEAP})-\log_{10}(X_{\rm lit})$, for the set of spectral parameters compared in each sample.
Panel~A shows against \citetalias{2022Wu}, while Panel~B shows against \citet{2025Pan}.
In each panel, the legend reports the compared parameters. The grey dashed line indicates the 1:1 relation, while the black dashed-lines demark the $\pm0.3$~dex.}
    \label{fig:RSP1}
\end{figure}

\begin{figure}
    \centering
    \includegraphics[width=.85\linewidth]{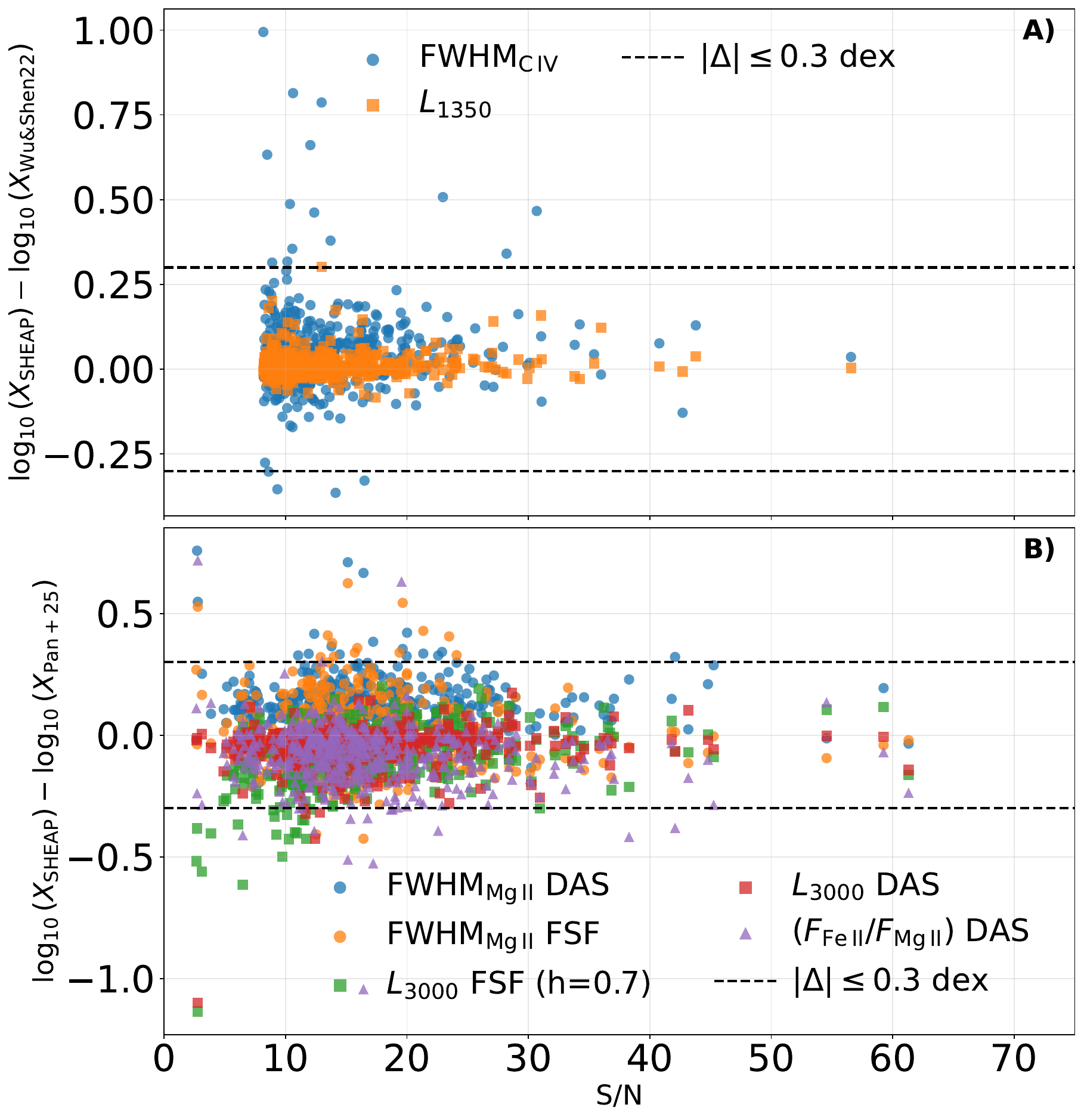}
   \caption{
Differences between our measurements and those from the literature as a function of signal-to-noise ratio (S/N; x-axis). The y-axis shows $\log_{10}(X_{\rm SHEAP})-\log_{10}(X_{\rm lit})$.
Panel~A against \citetalias{2022Wu}, and Panel~B against \citetalias{2025Pan}.
The legend identifies each parameter, and the horizontal dashed lines mark the $\pm0.3$ dex band.}
    \label{fig:Pair1SN}
\end{figure}

The full distributions of the logarithmic differences are shown in Fig.~\ref{fig:RSP1}A, and their dependence on S/N is shown in Fig.~\ref{fig:Pair1SN}A. Overall, we find good agreement with \citetalias{2022Wu}, with no strong increase in scatter toward lower S/N, although the largest outliers occur preferentially at the lowest S/N.

\subsection{The Pan+25 sample} 
The second test uses DESI DR1 spectra, which provide higher spectral resolution than SDSS and a uniform wavelength sampling across all spectra. We draw a subsample from the \textit{DESI DR1 value-added catalog v1.7} \citep[hereafter the \textit{BHMass Catalog};][]{2025Pan}, which contains spectral properties for $490{,}468$ objects over $0.6 \le z \le 1.6$ and covers the Mg\,\textsc{ii} region. The catalog reports measurements from two independent pipelines, \texttt{FastSpecFit} \citep[\texttt{FSF};][]{2023Moustakas} and \texttt{DASpec} \citep[\texttt{DAS};][]{2024Du}.

We select 500 sources over $0.60 \le z \le 0.622$ and download their spectra following the official DESI data access procedure.\footnote{\url{https://data.desi.lbl.gov/doc/access/}} Each spectrum contains 7{,}781 pixels, and the redshift and S/N distributions are shown in Fig.~\ref{fig:redshift_snr_fraction}. We compare our measurements with the catalog values of Mg\,\textsc{ii} FWHM, $L_{3000}$, and $R_{\mathrm{Fe\,II}}$, where
\begin{equation}
\label{eq:rfe}
    R_{\mathrm{Fe\,II}} = \frac{F_{\mathrm{Fe}}}{F_{\mathrm{Mg\,II}}},
\end{equation}
with $F_{\mathrm{Fe}}$ measured over 2{,}250--2{,}650~\AA.

Since this test focuses on the Mg\,\textsc{ii} region, we fit the spectra over 1{,}750--4{,}400~\AA. The model includes a power-law continuum, Fe\,\textsc{ii} emission, the Balmer continuum, higher-order Balmer lines, and one broad Gaussian component for the emission lines. Although \citetalias{2025Pan} model Mg\,\textsc{ii} with multiple broad and narrow components, we find that adding a second broad Gaussian does not significantly improve the fit quality. We therefore retain the simplest 26-parameter model for this validation test.

The fit-quality distribution gives a median $\chi^2_{\rm red}=1.3 \pm 0.25$, with more than 90\% of the objects having $\chi^2_{\rm red}\leq 2$ (Fig.~\ref{fig:SNchi2plots}). We observe a mild increase in $\chi^2_{\rm red}$ toward higher S/N, likely because higher-quality spectra reveal small model imperfections more clearly. Representative spectra close to the median S/N and median $\chi^2_{\rm red}$ are shown in Figs.~\ref{fig:spectra_pan_median_snr} and \ref{fig:spectra_pan_median_chi2}.

\begin{figure}
    \centering
    \includegraphics[width=0.65\linewidth]{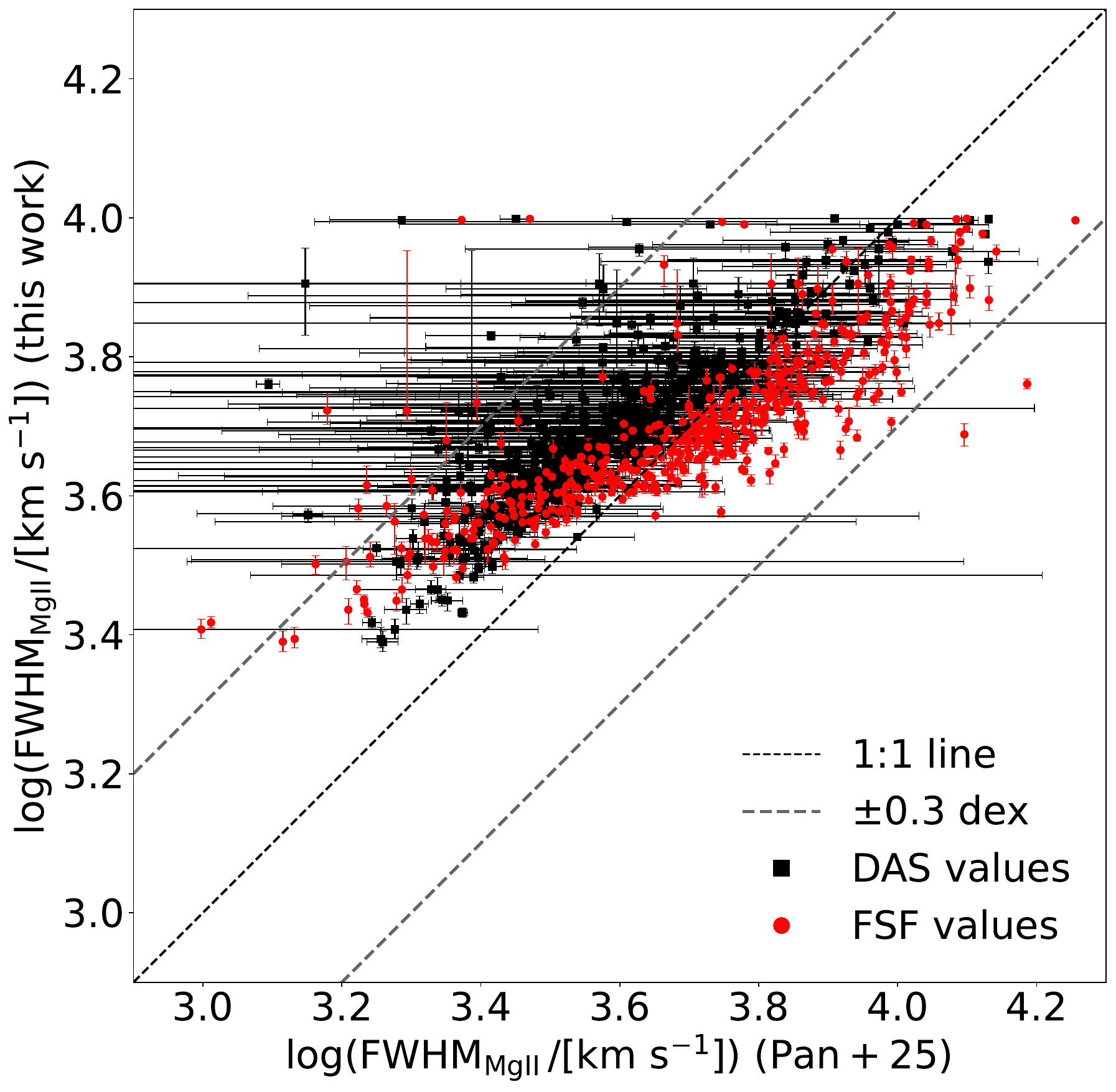}
    \vspace{0.25cm}
    \includegraphics[width=0.65\linewidth]{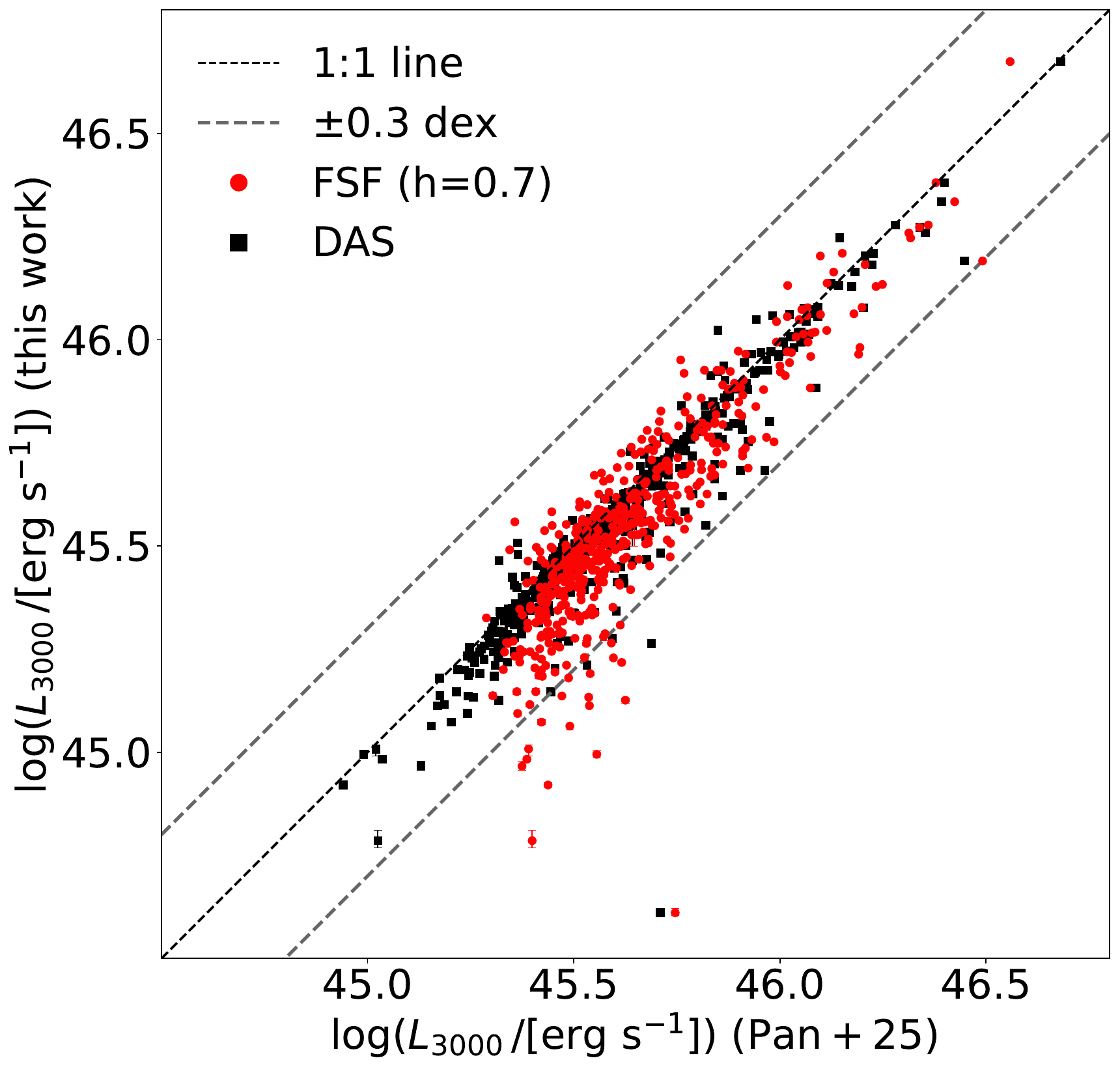}
    \caption{Comparison between our measurements and those of \citetalias{2025Pan}. Top: Mg\,\textsc{ii} emission-line FWHM in logarithmic scale. Bottom: Monochromatic continuum luminosity at 3000\,\AA\ in logarithmic scale. In both panels, the black dashed line indicates the 1:1 relation, while the gray lines mark the $\pm 0.3$ dex region. Red circles represent values computed with \texttt{FSF}, while black squares correspond to \texttt{DAS}.}
    \label{fig:2025PanComparison}
\end{figure}

The comparison with the catalog measurements is shown in Fig.~\ref{fig:2025PanComparison}. For $\mathrm{FWHM}_{\mathrm{Mg\,II}}$, 96.6\% of the \texttt{DAS} values and 97.0\% of the \texttt{FSF} values lie within the $\pm0.3$ dex band, with median biases of $+0.12$ and $-0.02$ dex, respectively. For $L_{3000}$, the agreement is tighter: 99.20\% of the \texttt{DAS} values and 96.00\% of the corrected \texttt{FSF} values lie within the band, with median biases of $-0.02$ and $-0.07$ dex, respectively. These results indicate that the continuum luminosity is robustly recovered, while the FWHM comparison is more sensitive to differences in the adopted Mg\,\textsc{ii} decomposition and Fe\,\textsc{ii} treatment.

Finally, we compare $R_{\mathrm{Fe\,II}}$ against the \texttt{DAS} measurements only. We find that 96.80\% of the objects lie within the $\pm0.3$ dex band, with a median bias of $-0.07$ dex and $\mathrm{NMAD}=0.01$ dex. The systematic underestimation of $R_{\mathrm{Fe\,II}}$ is consistent with the broader trends shown in Fig.~\ref{fig:RSP1}B, where our method generally returns slightly smaller values than the catalog, except for quantities directly tied to the Mg\,\textsc{ii} profile. As shown in Fig.~\ref{fig:Pair1SN}B, most outliers occur at low S/N, with no strong increase in scatter toward higher S/N.

\begin{figure}
    \centering
    \includegraphics[width=.75\linewidth]{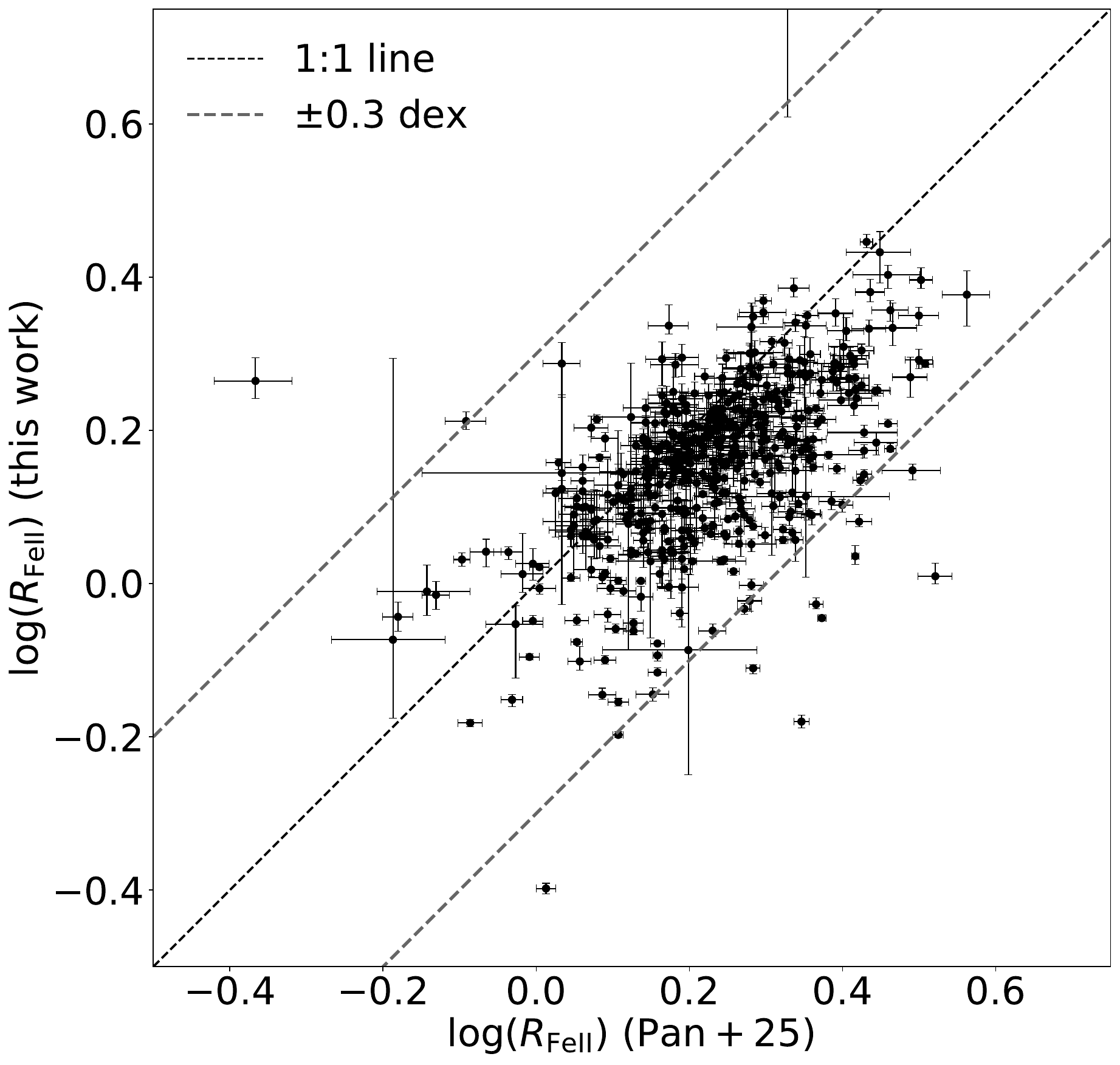}
     \caption{Comparison between our results and \cite{2025Pan} for the  $R_{\mathrm{Fe II}}$ in logarithmic scale. The black dashed line indicates the 1:1 relation, while the grey lines demark the $\pm0.3$~dex.}
    \label{fig:RfeII}
\end{figure}
\subsection{The Sánchez-Sáez+18 sample} 

In \citet{2018SanchezSaez} (hereafter, \citetalias{2018SanchezSaez}), the authors analyzed 2,345 SDSS AGN spectra using the methodology of \citet{2016MejiaRestrepo}. For 224 sources, they performed detailed fits to the H$\beta$ and Mg\,\textsc{ii} regions, including host-galaxy subtraction through the \textit{k-star} method \citep{2005Greene,2006Kim}. From this set, we retain 151 objects with $0.45 \le z \le 0.8$, which provides rest-frame coverage from 3,000 to 5,000~\AA\ and enables a simultaneous analysis of the H$\beta$ and Mg\,\textsc{ii} regions. The S/N and redshift distributions are shown in Fig.~\ref{fig:redshift_snr_fraction}. Since this sample is based on SDSS spectra, we resample it as described in Section~\ref{sec:WuandShen}.

We compare our results with the reported FWHM values for H$\beta$ and Mg\,\textsc{ii}, together with the monochromatic continuum luminosities at 5{,}100 and 3{,}000~\AA. The reference measurements were obtained by modeling the spectra with a double-Gaussian description of the broad emission lines, a narrow component, and Fe\,\textsc{ii} emission, after subtracting the host-galaxy contribution based on the Ca\,\textsc{ii}\,K absorption feature. In our analysis, we fit the common 2{,}450--5{,}500~\AA\ window using two broad components plus a narrow component, Fe\,\textsc{ii} emission, Balmer continuum and higher-order Balmer lines, possible outflow components, a power-law continuum, and our host-galaxy modeling method (Section~\ref{sec:host}). The resulting model contains 421 parameters.

This sample has the lowest S/N distribution among our tests. Nevertheless, the fit quality remains good, with a median $\chi^2_{\rm red}=1.2 \pm 0.2$ and no objects exceeding $\chi^2_{\rm red}\sim3$ (Fig.~\ref{fig:SNchi2plots}). As in the previous samples, $\chi^2_{\rm red}$ shows a mild increase with S/N, while low-S/N spectra remain close to unity. Representative fits near the median S/N and median $\chi^2_{\rm red}$ are shown in Figs.~\ref{fig:spectra_SS_median_snr} and \ref{fig:spectra_SS_median_chi2}. Even in the lowest-S/N cases, the main components are recovered, although Fe\,\textsc{ii} is less tightly constrained.

\begin{figure}
    \centering
    \includegraphics[width=0.65\linewidth]{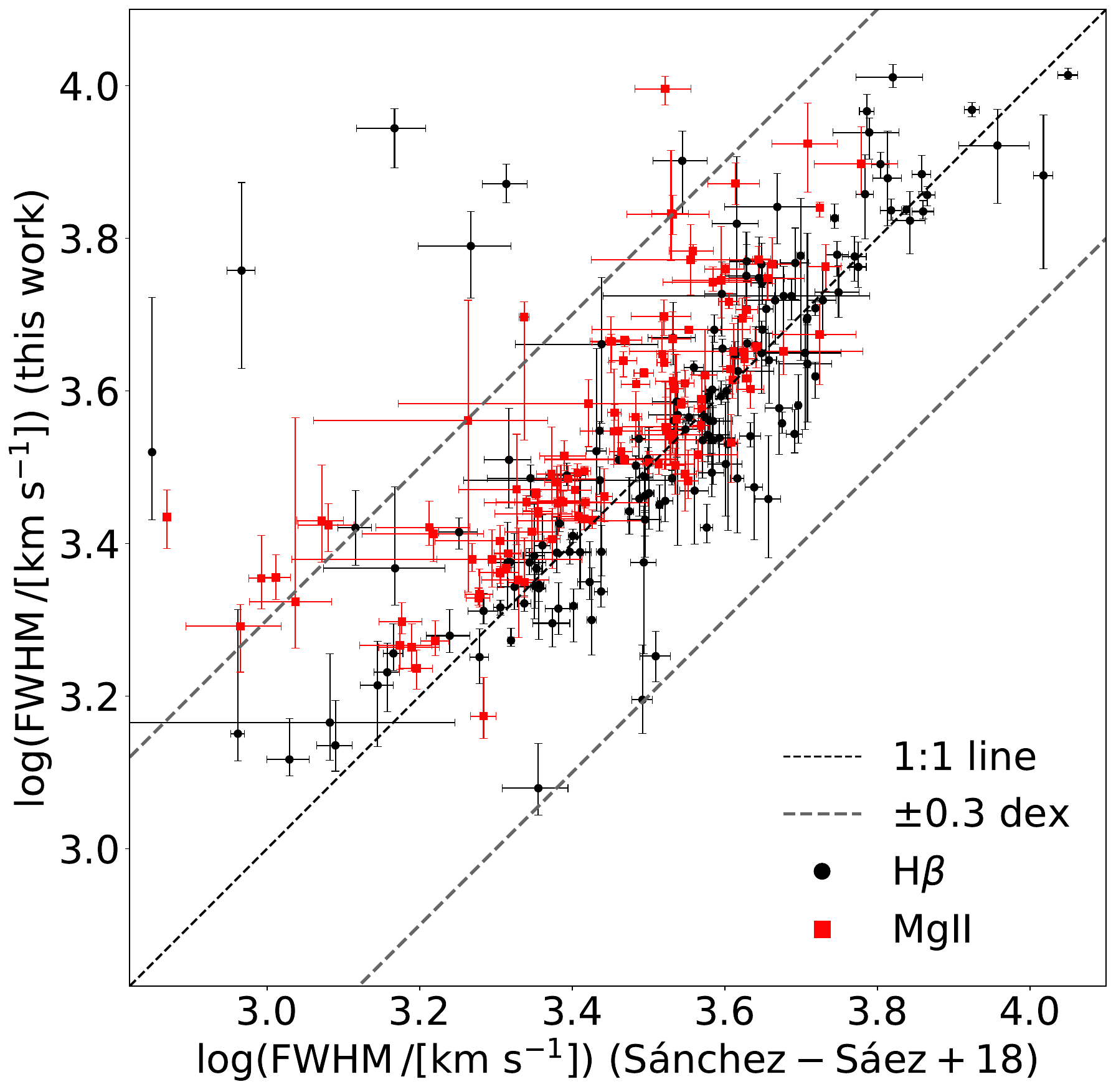}
    \vspace{0.25cm}
    \includegraphics[width=0.65\linewidth]{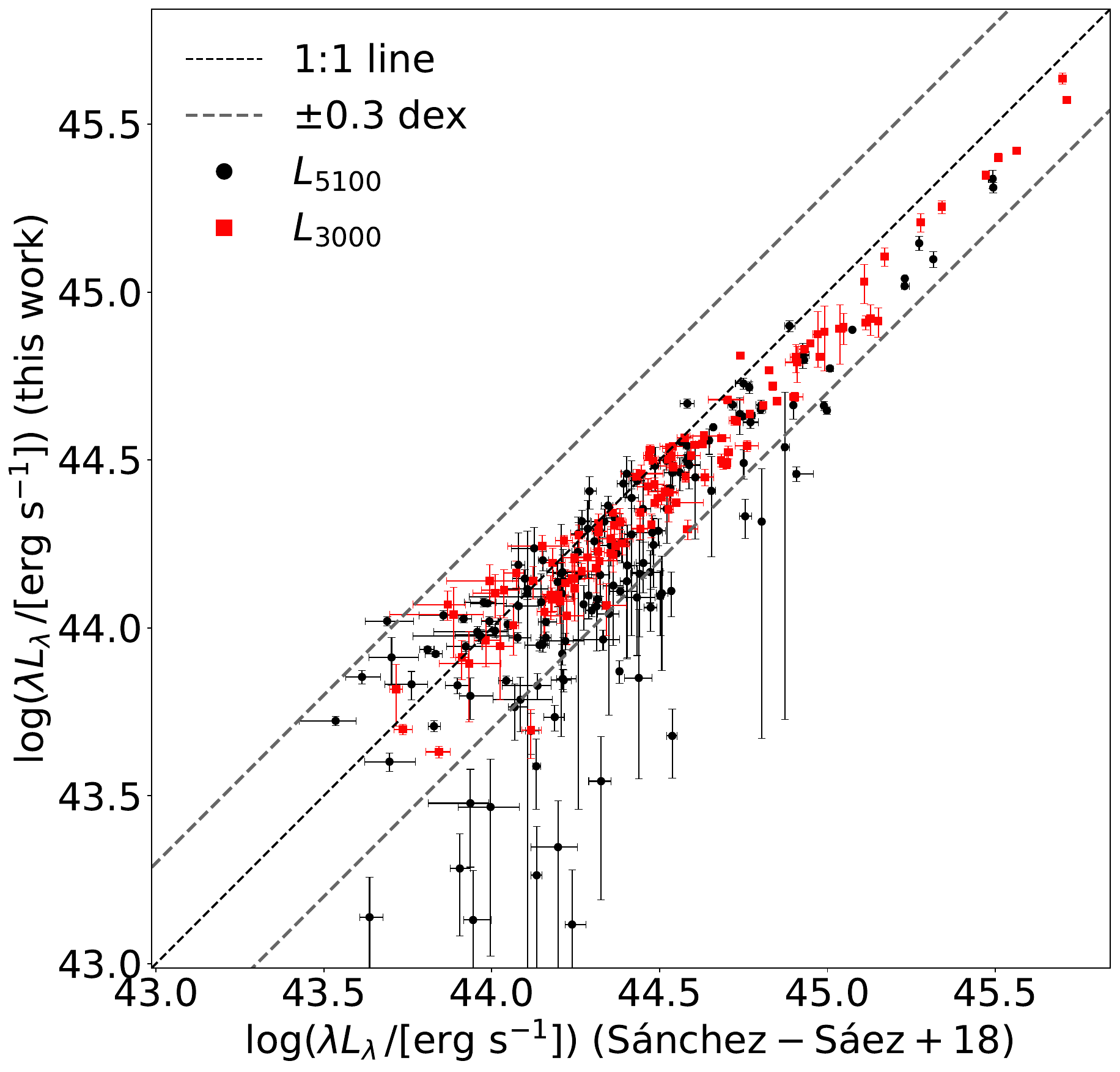}
    \caption{Comparison between our measurements and those of \citetalias{2018SanchezSaez}. Top: comparison of the emission-line FWHM in logarithmic scale. Bottom: comparison of the monochromatic continuum luminosity in logarithmic scale. In both panels, the black dashed line indicates the 1:1 relation, while the gray lines mark the $\pm 0.3$ dex region. In the top panel, black circles correspond to H$\beta$ measurements, while red squares represent Mg\,\textsc{ii}. In the bottom panel, black circles correspond to $L_{5100}$ measurements, while red squares represent $L_{3000}$.}
    \label{fig:2018SSComparison}
\end{figure}

Line widths are measured from the combined broad components using the ``combining all broad components'' approach (Section~\ref{sec:combiningall}). The comparison is shown in Fig.~\ref{fig:2018SSComparison}Top. For $\mathrm{FWHM}_{\rm H\beta}$, 95.36\% of the objects fall within the $\pm0.3$ dex band, with a median bias of $+0.01$ dex and $\mathrm{NMAD}=0.08$ dex. For $\mathrm{FWHM}_{\rm Mg\,II}$, considering the 117 objects with available catalog values, 92.30\% lie within the band, with a median bias of $+0.08$ dex and $\mathrm{NMAD}=0.08$ dex. The somewhat larger Mg\,\textsc{ii} values may reflect the degeneracy between Fe\,\textsc{ii} emission and the Mg\,\textsc{ii} wings.

The continuum-luminosity comparison is shown in Fig.~\ref{fig:2018SSComparison}Bottom. For $L_{5100}$, 77.48\% of the objects lie within the $\pm0.3$ dex band, with a median bias of $-0.13$ dex and $\mathrm{NMAD}=0.17$ dex. This offset is likely driven by differences in host-galaxy subtraction, since the \textit{k-star} method relies on the Ca\,\textsc{ii}\,K absorption feature, whereas our approach uses a broader set of spectral features. For $L_{3000}$, the agreement is stronger: 99.12\% of the 117 objects with available catalog values fall within the band, with a median bias of $-0.10$ dex and $\mathrm{NMAD}=0.07$ dex.

\begin{figure}
    \centering
    \includegraphics[width=.85\linewidth]{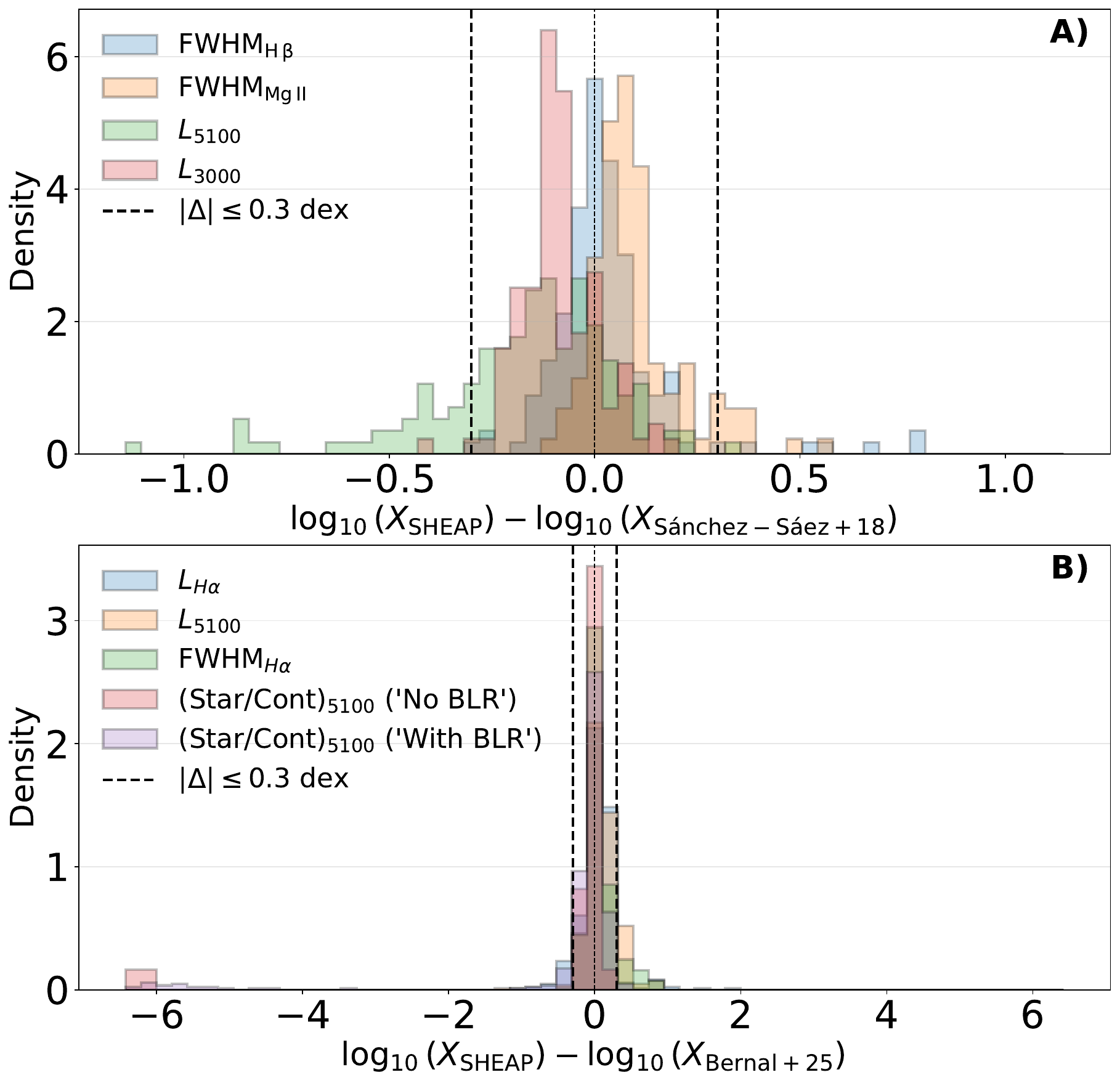}
  \caption{Same as Fig.~\ref{fig:RSP1} but this time Panel~A shows the comparison against \citetalias{2018SanchezSaez}, meanwhile Panel~B shows the comparison against \citetalias{2025Bernal}.}
    \label{fig:RSP2}
\end{figure}

\begin{figure}
    \centering
    \includegraphics[width=.85\linewidth]{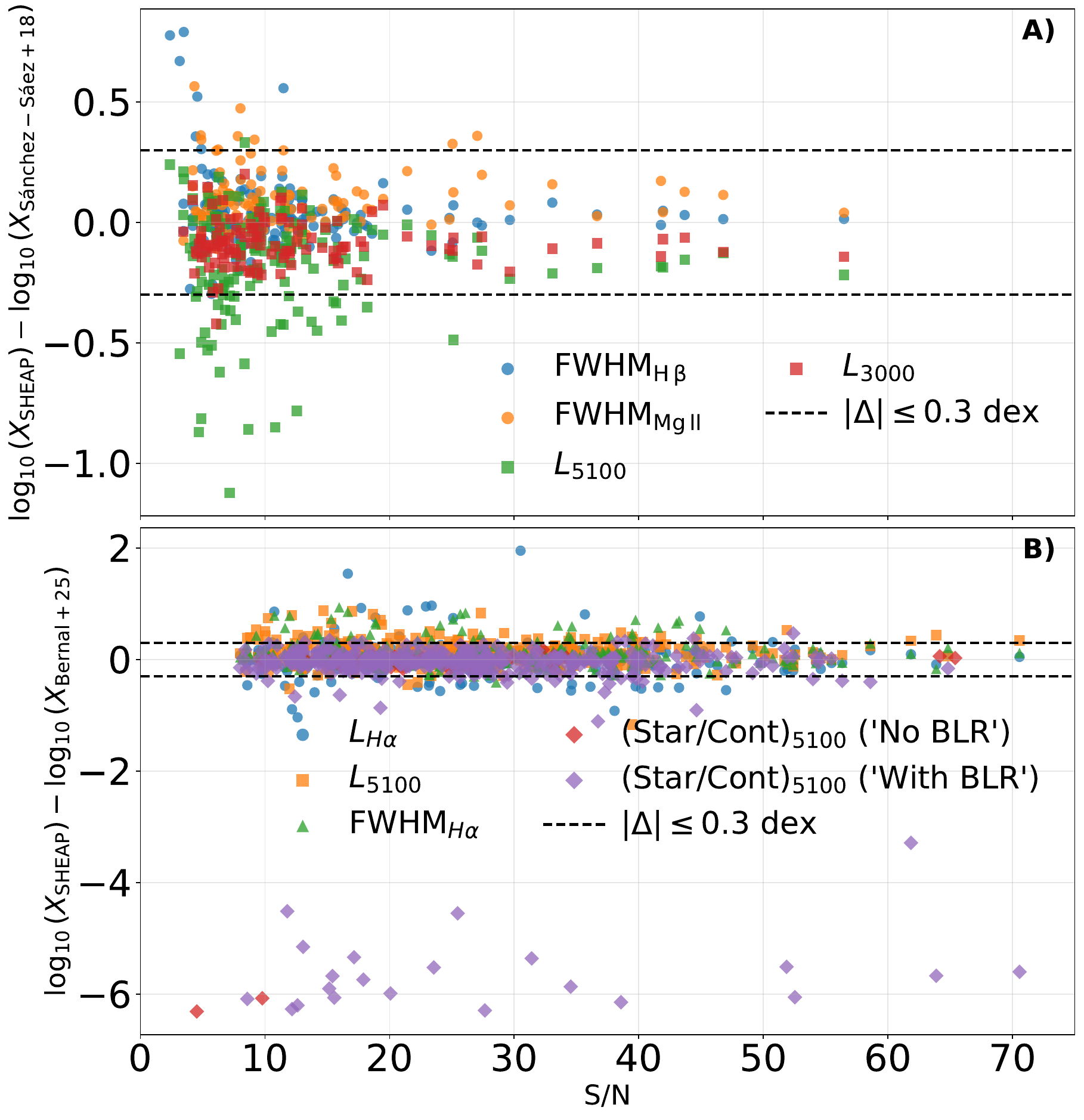}
\caption{Same as Fig.~\ref{fig:Pair1SN} but this time Panel~A compares against \citetalias{2018SanchezSaez}, and Panel~B compares against \citetalias{2025Bernal}.}

    \label{fig:Pair2SN}
\end{figure}

Overall, the parameter-difference distributions are shown in Fig.~\ref{fig:RSP2}A, and their dependence on S/N is shown in Fig.~\ref{fig:Pair2SN}A. The largest discrepancies occur preferentially at low S/N, but we do not find evidence for a strong systematic trend with S/N.

\subsection{The Bernal+25 sample} 

To evaluate the performance of our method in a low-redshift, host-dominated sample, we use the dataset presented by \citetalias{2025Bernal}. This sample spans $0.01 \le z \le 0.14$ and was analyzed with a customized \texttt{pPXF} wrapper to decompose the AGN and host-galaxy components. From the publicly available catalog,\footnote{\url{https://cdsarc.cds.unistra.fr/viz-bin/cat/J/A+A/694/A127}} we construct a sample of 413 objects, whose redshift and S/N distributions are shown in Fig.~\ref{fig:redshift_snr_fraction}. The spectra cover 3,559--10,394~\AA\ in the observed frame and, since they are based on SDSS data products, are resampled as described in Section~\ref{sec:WuandShen}.

We compare the host-to-continuum ratio at 5{,}100~\AA, $(\mathrm{Star/Cont})_{5100}$, the monochromatic luminosity $L_{5100}$, the FWHM of the broad H$\alpha$ component, and the broad H$\alpha$ luminosity. To obtain comparable measurements, we adopt a model similar to that of \citetalias{2025Bernal}, including two broad components, a narrow component, Fe\,\textsc{ii} emission, Balmer emission, and a host-galaxy component. We fit the rest-frame wavelength range 3,800--8,000~\AA, resulting in a model with 427 parameters.

Despite the complexity of these spectra, the reduced chi-square distribution has a median of $\chi^2_{\rm red}=1.8 \pm 0.9$ (Fig.~\ref{fig:SNchi2plots}). As in the other samples, $\chi^2_{\rm red}$ increases mildly with S/N, likely because higher-S/N spectra reveal small model imperfections more clearly. A representative fit near the median S/N is shown in Fig.~\ref{fig:spectra_bernal_median_snr}, where the host-galaxy component is well reproduced.

\begin{figure}
    \centering
    \includegraphics[width=0.65\linewidth]{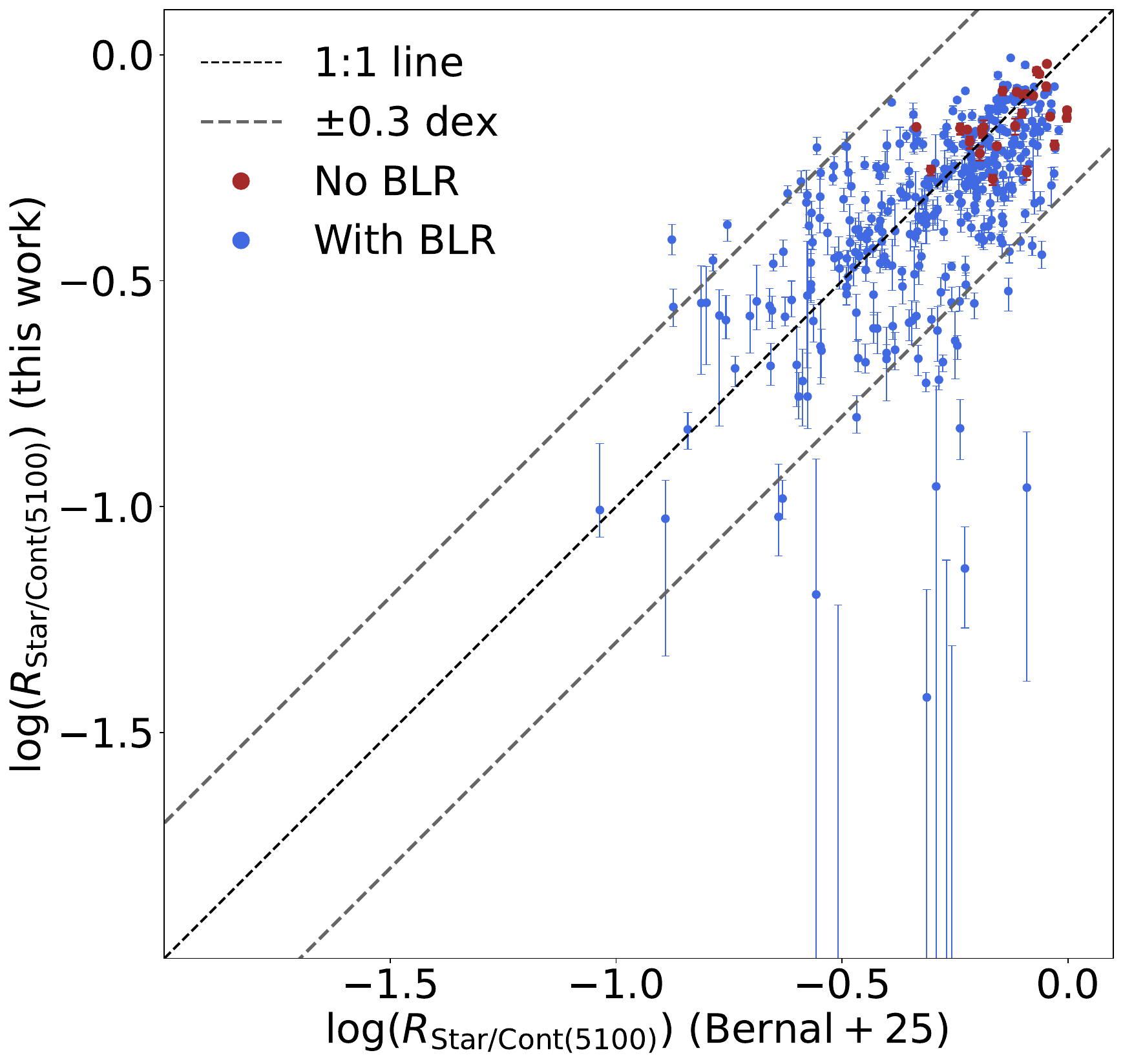}
    \caption{Comparison between the stellar-to-continuum flux ratio at 5{,}100 \AA\,derived by \citetalias{2025Bernal} and the values obtained in this work. Red points show all objects in the sample, while blue points highlight sources with a confirmed broad emission-line component. The dashed line marks the 1:1 relation, and the shaded region indicates the $\pm$0.3 dex agreement band.}
    \label{fig:2025Bernalscont5100}
\end{figure}

For $(\mathrm{Star/Cont})_{5100}$, 86.90\% of the objects lie within the $\pm0.3$ dex band, with a median bias of $-0.04$ dex and $\mathrm{NMAD}=0.129$ dex (Fig.~\ref{fig:2025Bernalscont5100}). When separating the sample according to the BLR classification of \citetalias{2025Bernal}, the agreement is 86.42\% for the BLR subsample and 93.10\% for the no-BLR subsample. A small number of extreme outliers are present, but visual inspection shows that most BLR outliers do not exhibit a clear host-galaxy contribution, while the no-BLR outliers have unreliable fits with $\chi^2_{\rm red}>10$.

\begin{figure}
    \centering
    \includegraphics[width=0.65\linewidth]{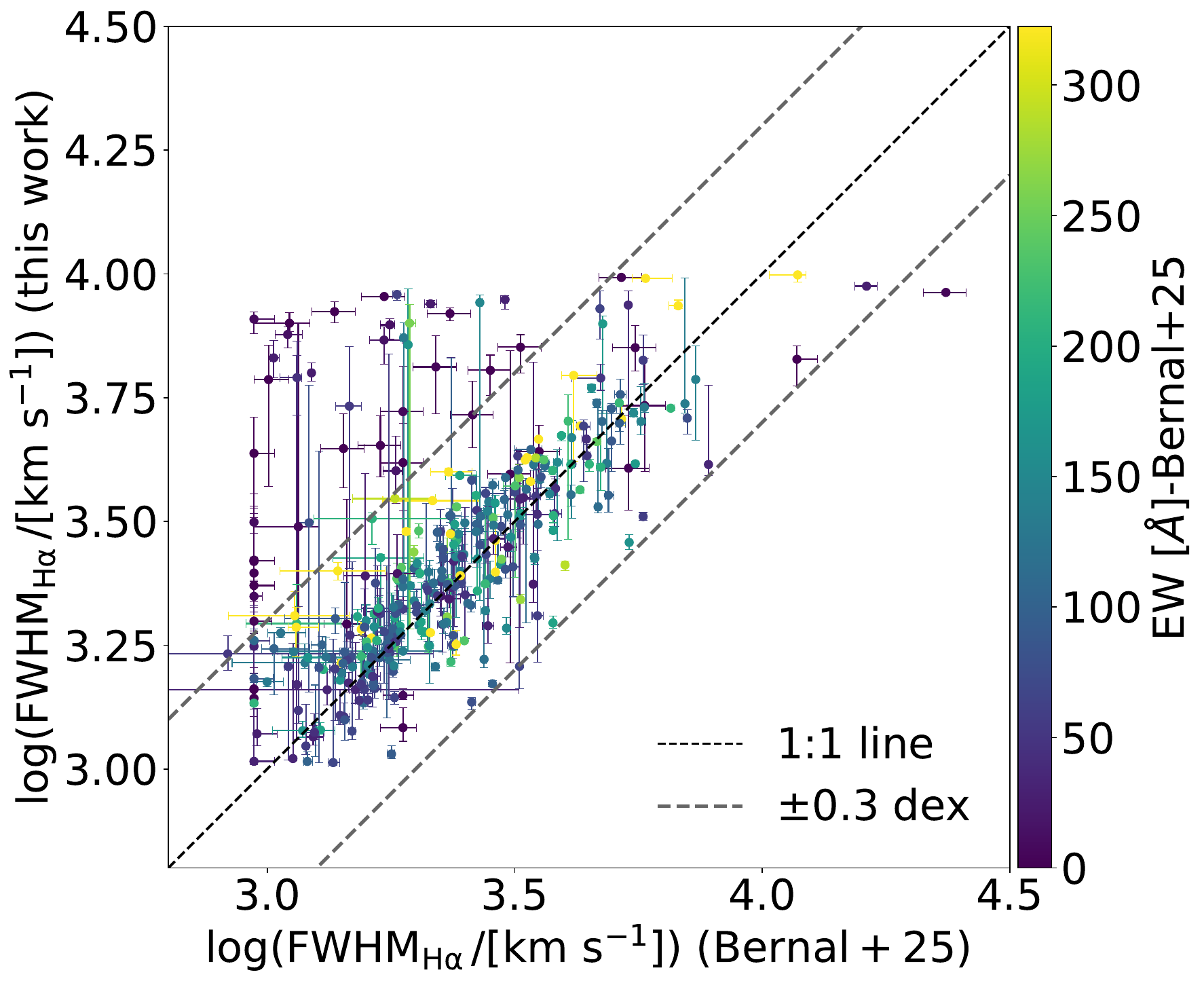}
    \vspace{0.25cm}
    \includegraphics[width=0.65\linewidth]{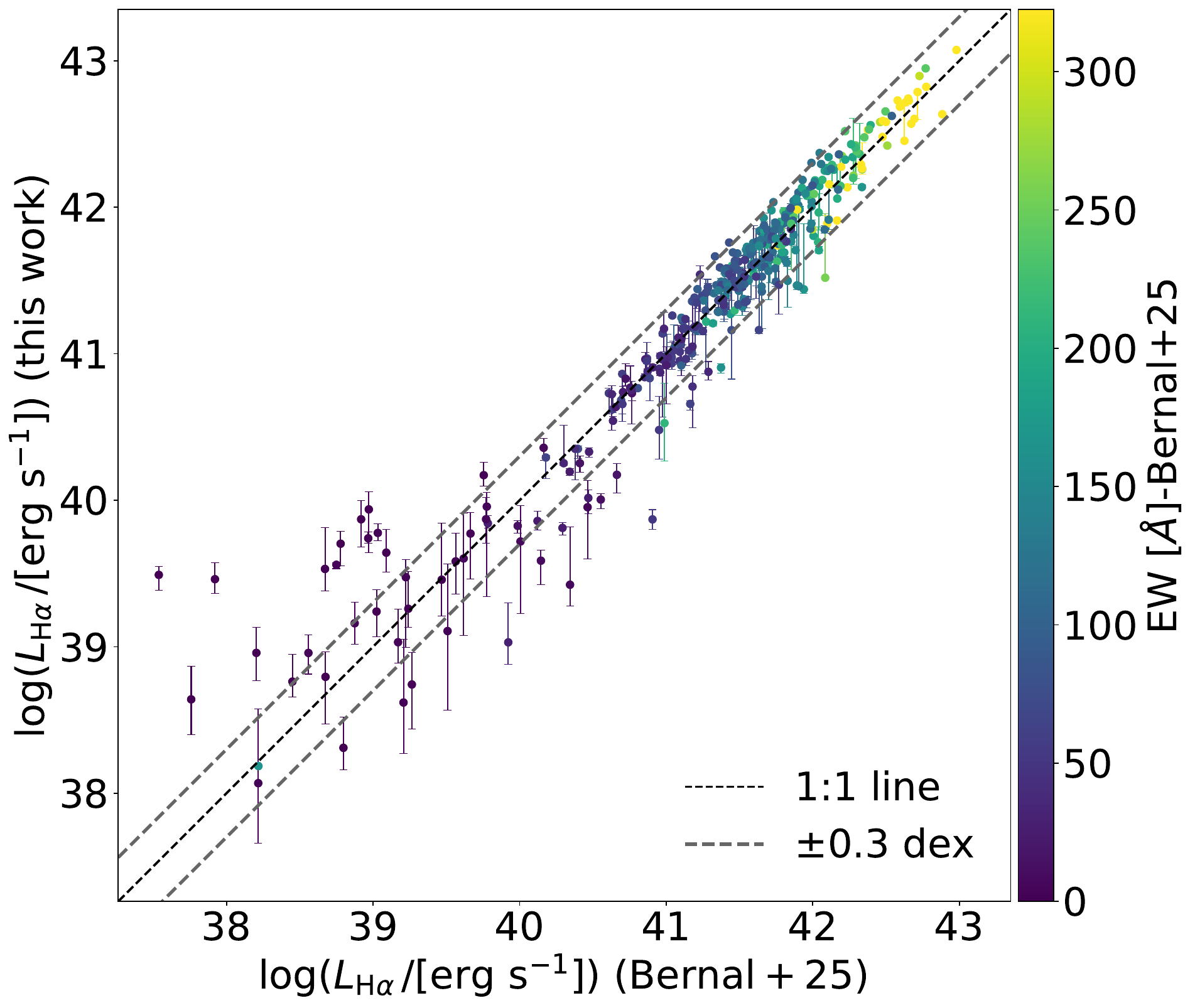}
    \caption{Comparison between our measurements and those of \citetalias{2025Bernal}. Top: H$\alpha$ emission-line FWHM in logarithmic scale. Bottom: H$\alpha$ emission-line luminosity in logarithmic scale. In both panels, the black dashed line indicates the 1:1 relation, while the gray lines mark the $\pm 0.3$ dex region. Point colors represent the EW reported by \citetalias{2025Bernal}.}
    \label{fig:2025BernalHalphaComparison}
\end{figure}

For the remaining comparisons, we restrict the analysis to the 384 objects with a detected broad component. The H$\alpha$ FWHM and luminosity comparisons are shown in Fig.~\ref{fig:2025BernalHalphaComparison}. For $\mathrm{FWHM}_{\mathrm{H}\alpha}$, using the kinematic-combination method (Section~\ref{sec:combiningkine}), 88.80\% of the objects fall within the $\pm0.3$ dex band, with a median bias of $+0.04$ dex and $\mathrm{NMAD}=0.11$ dex. For the broad H$\alpha$ luminosity, 87.76\% fall within the band, with a median bias of $+0.04$ dex and $\mathrm{NMAD}=0.15$ dex. In both cases, the largest discrepancies occur preferentially in low-EW systems, while objects with ${\rm EW}\gtrsim100$~\AA\ show the tightest agreement.

\begin{figure}
    \centering
    \includegraphics[width=.65\linewidth]{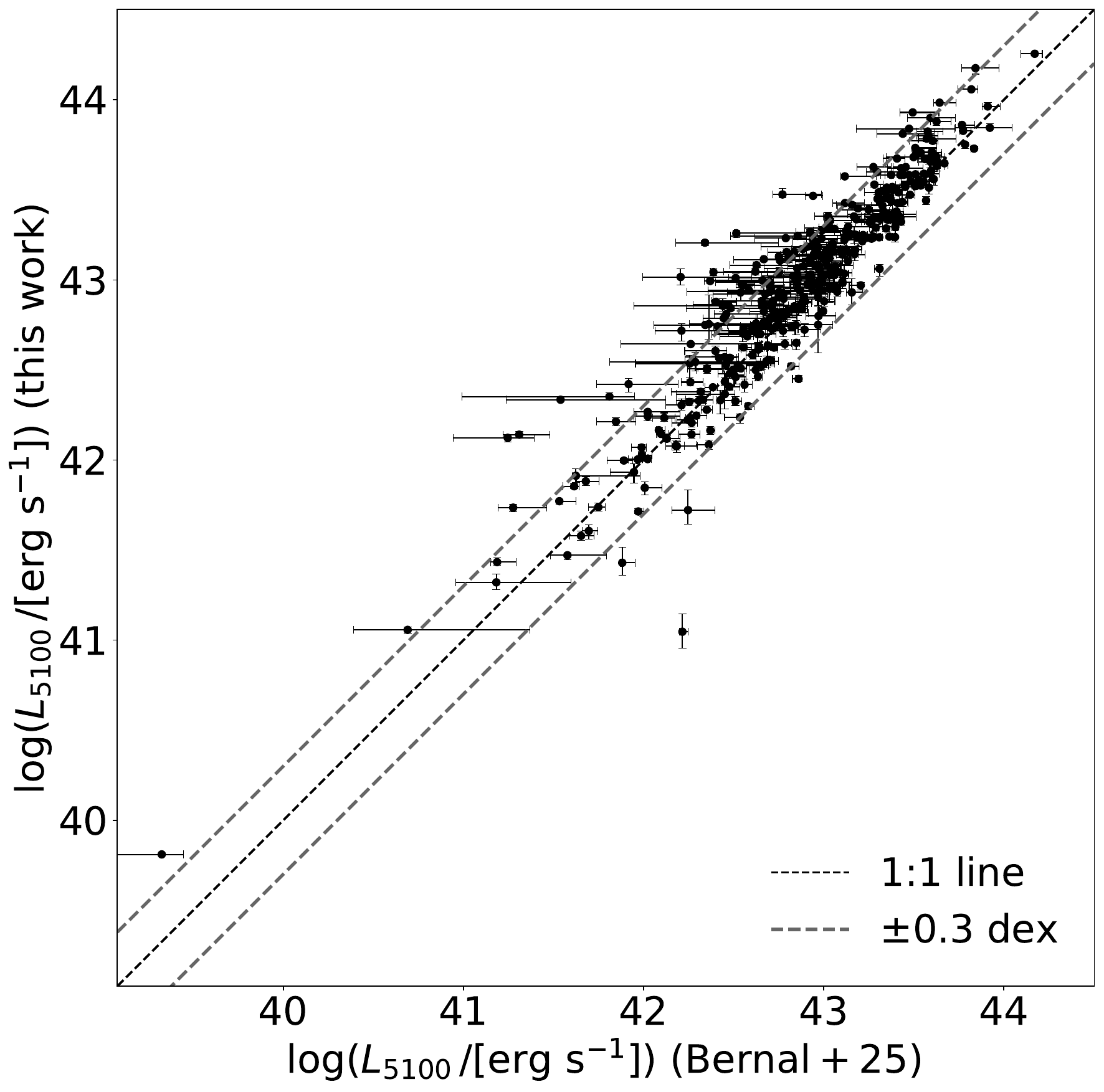}
    \caption{Comparison between our results and \citetalias{2025Bernal} for the monochromatic continuum luminosity at 5{,}100\,\AA\, in log scale.The black dashed line indicates the 1:1 relation, while the grey lines demark the $\pm0.3$~dex.}

    \label{fig:2025BernalL5100}
\end{figure}

Finally, for $L_{5100}$, 84.33\% of the objects lie within the $\pm0.3$ dex band, with a median bias of $+0.08$ dex and $\mathrm{NMAD}=0.15$ dex (Fig.~\ref{fig:2025BernalL5100}). The parameter-difference distribution in Fig.~\ref{fig:RSP2}B shows that our values are systematically larger, consistent with differences in the AGN--host decomposition: relative to \citetalias{2025Bernal}, our fits assign a smaller contribution to the stellar component, increasing the inferred AGN continuum luminosity at 5{,}100~\AA.

\begin{table*}
\centering
\caption{Summary of fit quality, model complexity, and computational performance for the literature comparison samples.}
\label{tab:chi2_times}
\begin{tabular}{lccccccc}
\hline
Sample & $N_{\rm spec}$ & $\tilde{\rm S/N}$ & $N_{\lambda}$  & $N_{\rm par}$ & $\tilde{\chi}^2_{\rm red}$ & $t_{\rm fit}$ [min] & $t_{\rm MC}$ [min] \\
\hline
Wu\&Shen+22 (C\,\textsc{iv}) & 500 & 11.93 & 3809  & 41 & $4.3^{+5.5}_{-1.8} (1.5^{+1.4}_{-0.5})$ & 0.2 & 5.3 \\[2pt]
Pan+25 (Mg\,\textsc{ii}) & 500 & 15.25 & 7781  & 26 & $1.3^{+0.4}_{-0.1}$ & 0.9 & 42.1 \\[2pt]
S\'anchez-S\'aez+18 (H$\beta$/Mg\,\textsc{ii}) & 151 & 9.33 & 3808  & 421 & $1.2^{+0.3}_{-0.1}$ & 0.9 & 16.9 \\[2pt]
Bernal+25 (H$\alpha$) & 413 & 24.13 & 2878  & 427 & $1.8^{+1.4}_{-0.4}$ & 1.3 & 35.5 \\[2pt]
\hline
\cite{2026Bernal} (\texttt{pPXF} iteration) & 413 &-- & -- & -- & -- & $130.8^{+123.9}_{-41.3}$ & -- \\[2pt]
\hline
\end{tabular}
\tablefoot{
$\widetilde{\chi}^2_{\rm red}$ denotes the median reduced chi-square, reported together with the 16th and 84th percentiles as asymmetric uncertainties.
For Wu\&Shen+22, we report the unmasked value, with the value obtained after masking $\lambda < 1300$\,\AA\ given in parentheses.
$\widetilde{\rm S/N}$ is the median signal-to-noise ratio of the sample.
$N_{\rm par}$ is the number of free parameters used in the fit, corresponding to the length of the optimized parameter vector.
$N_{\lambda}$ denotes the number of pixels per spectrum.
$t_{\rm fit}$ is the runtime for a single best-fit optimization per spectrum, and $t_{\rm MC}$ is the runtime for the Monte Carlo posterior stage per spectrum.
}
\end{table*}

\section{Discussion \& Conclusions}
\label{sec:Discussion}

In this work, we presented \texttt{SHEAP}, a GPU-enabled AGN spectral-fitting code designed for efficient, flexible, and scalable analysis of large spectroscopic datasets. We validated its performance using four literature-based comparison samples spanning different redshift ranges, spectral regions, resolutions, and AGN populations, including SDSS and DESI spectra covering the C\,\textsc{iv}, Mg\,\textsc{ii}, H$\beta$, and H$\alpha$ regions. This strategy demonstrates that \texttt{SHEAP} performs robustly under heterogeneous observational and modelling conditions, rather than being tuned to a single homogeneous dataset.

Overall, \texttt{SHEAP} provides statistically acceptable fits across all samples. The reduced chi-square distributions are typically centered near unity, with median values of $\tilde{\chi}^2_{\rm red}\sim1.2$--$1.8$ for the Pan+25, S\'anchez-S\'aez+18, and Bernal+25 samples. The Wu\&Shen+22 sample initially shows a broader distribution, but after masking the region at $\lambda_{\rm rest}<1300$~\AA, affected by the Ly$\alpha$ forest, the median $\tilde{\chi}^2_{\rm red}$ improves from $4.3^{+5.5}_{-1.8}$ to $1.5^{+1.4}_{-0.5}$. Across all samples, we do not find evidence for a strong degradation of fit quality toward lower S/N; instead, the mild increase of $\chi^2_{\rm red}$ at high S/N likely reflects the increased visibility of subtle spectral features and modelling imperfections.

The comparison with literature measurements shows that \texttt{SHEAP} recovers the main AGN spectral parameters with generally good agreement. The fraction of objects within the $\pm0.3$ dex band is typically above 90\% for line-width and continuum-luminosity measurements in the Wu\&Shen+22, Pan+25, and S\'anchez-S\'aez+18 samples, including $\mathrm{FWHM}_{\mathrm{C\,IV}}$, $\mathrm{FWHM}_{\mathrm{Mg\,II}}$, $\mathrm{FWHM}_{\mathrm{H}\beta}$, $L_{1350}$, $L_{3000}$, and $R_{\mathrm{Fe\,II}}$. Lower agreement is found for host-sensitive quantities, such as $L_{5100}$ and $(\mathrm{Star/Cont})_{5100}$, particularly in the S\'anchez-S\'aez+18 and Bernal+25 samples, where differences in host-galaxy subtraction and AGN--host decomposition become more important.

The main residual discrepancies are therefore linked to methodological differences rather than to a single systematic bias in \texttt{SHEAP}. In the UV and Mg\,\textsc{ii} regions, differences are mainly associated with the parameterization and combination of broad-line components, as well as the treatment of Fe\,\textsc{ii} emission. In the optical region, the largest offsets arise in continuum-related quantities, especially $L_{5100}$, where the balance between stellar and AGN continuum components becomes critical.

A major result of this work is that this level of agreement is achieved at low computational cost. The runtime depends more strongly on the number of pixels per spectrum and the number of spectra being fitted than on the formal number of free parameters, allowing models with hundreds of parameters to remain tractable. In absolute terms, the best-fit optimization requires between 0.2 and 1.3 minutes for the tested subsamples, whereas the Monte Carlo stage requires between 5.3 and 42.1 minutes per tested subsample, this including 50 Monte Carlo realizations plus the initial best-fit run.

As a direct benchmark, our implementation requires 1.3 minutes for the fitting stage, compared to the $130.8^{+123.9}_{-41.3}$ minutes reported by \cite{2026Bernal} for their \texttt{pPXF}-based iterative pipeline. This corresponds to an improvement factor of approximately 101 times, or $\sim1.0\%$ of the reported fitting time. This comparison should be interpreted with caution, since differences in hardware and computational environment may also contribute to the runtime difference.

Taken together, these results show that \texttt{SHEAP} can recover key AGN spectral properties across heterogeneous datasets while substantially reducing computational cost. Its GPU-accelerated, gradient-based framework makes it well suited for upcoming large spectroscopic surveys, where scalability, reproducibility, and flexible model configurations are essential. Future developments will extend this framework through more flexible line-profile parameterizations, applications to integral-field spectroscopy data cubes, and adaptations to related problems such as galaxy stellar-population analysis and dynamical spectral modelling.

\section*{Data availability}
In the body of the paper, we presented Figures and results for a series of comparisons. We include an online appendix that reproduces all the results presented in this work. We provide an online appendix at \url{https://github.com/felavila/Results_paper_SHEAP}.
\begin{acknowledgements}
      FAV acknowledges funding from the Doctorate Fellowship Program FIB-UV of the Universidad de Valparaíso. 
    FAV and PSS acknowledge support from the ESO Early-Career Visitor Programme and the ESO Science Support Discretionary Fund, project No. 4/25-G. 
    This work was supported by ANID FONDECYT Regular grant No. 1231418 (VM, FAV) and by the Centro de Astrofísica de Valparaíso (CIDI 21). 
    SB acknowledges support from the National Agency for Research and Development (ANID) through grant Gemini-32240014.

This research made use of Astropy, SpectRes, JAX, Matplotlib, NumPy, SciPy, pandas, Optax, uncertainties, and sfdmap2.
\end{acknowledgements}

%
%
\bibliographystyle{aa} 
  \bibliography{ref.bib} 
\onecolumn
\begin{appendix}

\section{Kinematic region limits}

\begin{table*}
\centering
\caption{\texttt{SHEAP} kinematic region limits.}
\label{tab:region_limits}
\renewcommand{\arraystretch}{1.15}
\begin{tabular}{lccc}
\hline\hline
Region &
FWHM limits &
$\pm v_{\rm shift}$ &
References \\
&
\multicolumn{1}{c}{[km\,s$^{-1}$]} &
\multicolumn{1}{c}{[km\,s$^{-1}$]} &
\\
\hline
Broad  & 1500--10000  & 3000 & \citealt{2011Shen,2017Rakshit} \\
Narrow & 100--500     & 500  & \citealt{2013Mullaney,2017Calderone} \\
Outflow & 500--1500   & 500  & \citealt{2013Mullaney,2018Circosta,2018Harrison} \\
Winds  & 3000--15000  & 8000 & \citealt{2011Richards,2016Coatman,2017Coatman} \\
Host   & 100--1000    & 2896 & \citealt{2003Bruzual,2006SanchezBlazquez,2004Cappellari} \\
Fe     & 100--7000    & 3000 & \citealt{2001Vestergaard,2010Kovacevic} \\
Balmer & 100--11000   & 500  & \citealt{2012Jin,2014Kovacevic} \\
\hline
\end{tabular}
\tablefoot{
Columns list the region name, the allowed FWHM range, the maximum absolute velocity offset, and the references used to define or motivate these limits.
}
\end{table*}
\twocolumn
\section{Broad Component Combination Methods}
\label{sec:broadcombination}
Broad emission lines in Type~1 AGNs are often modeled with multiple components (e.g., multiple Gaussians or others profiles, see Section~\ref{sec:EmissionLinesModels}) to reproduce asymmetric wings, intermediate-width structure, or blended sub-features. While this improves fit flexibility, it introduces the practical question of how to report a single set of broad-line measurements (e.g., velocity shift, line luminosity, and FWHM) for subsequent analysis.

In \texttt{SHEAP}, broad-line observables can be measured on reconstructed broad-line profiles obtained under different component-combination schemes. In particular, the FWHM is measured as the full width of a reconstructed broad-line profile at half the maximum of that profile peak. Because multi-component broad lines admit more than one physically motivated definition of the ``main'' broad profile, \texttt{SHEAP} computes broad-line measurements for two complementary strategies: (i) combining all fitted broad components, and (ii) combining only those broad components that are kinematically consistent with the systemic reference frame (as defined by the narrow lines). For the analyses presented in this work, we calculate both sets of measurements and compare them as appropriate to the sample under study; ultimately, the choice depends on the scientific application.

\paragraph{Combining all broad components.}
\label{sec:combiningall}
This method assumes that the set of broad components included in the model is an adequate description of the physical broad-line region signal. All broad components associated with a given transition are added, and a single set of broad-line observables is measured from the resulting total profile. This approach mirrors common practice in automated quasar-fitting tools, in which multiple broad subcomponents are treated collectively when reporting global broad-line properties \citep[e.g.,][]{2018Guo}.

\paragraph{Combining kinematically reasonable components}
\label{sec:combiningkine}
This alternative method does not assume that all fitted broad components are physically associated with the main virialized broad-line emission. Instead, it uses the systemic reference frame defined by narrow emission lines to determine which broad components are plausibly associated with the AGN systemic velocity. Narrow lines are widely used as kinematic anchors in AGN spectral decomposition and quasar catalogs \citep[e.g.,][]{2011Shen}. In \texttt{SHEAP}, a broad component is included in the ``main'' broad profile only if its velocity shift relative to the narrow-line systemic velocity falls within a predefined threshold ($|v_{\rm shift}| \le v_{\rm th}$).

In our implementation we adopt, for example, $v_{\rm th}=150~\mathrm{km~s^{-1}}$ following the criterion used in \citep[][]{2026Bernal,2025Bernal}. The combined profile is then formed by summing only the selected components, and all reported broad-line measurements (including FWHM, centroid shift, and integrated flux) are derived from that kinematically filtered total profile.

\section{Example modeled spectra}
\begin{figure}[!htbp]
    \centering
    \includegraphics[width=1.0\linewidth]{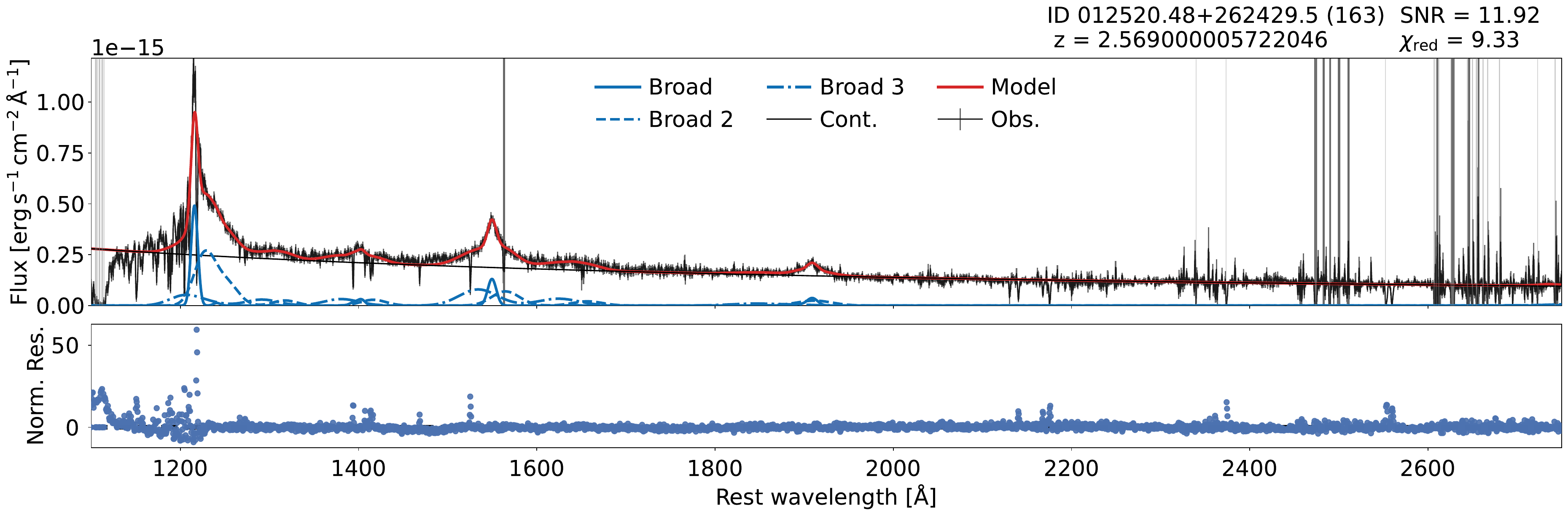}
    \caption{Example of fited spectra for the Wu\&Shen sample for the object with median S/N}
    \label{fig:spectra_wushen_median_snr}
\end{figure}

\begin{figure}[!htbp]
    \centering
    \includegraphics[width=1.0\linewidth]{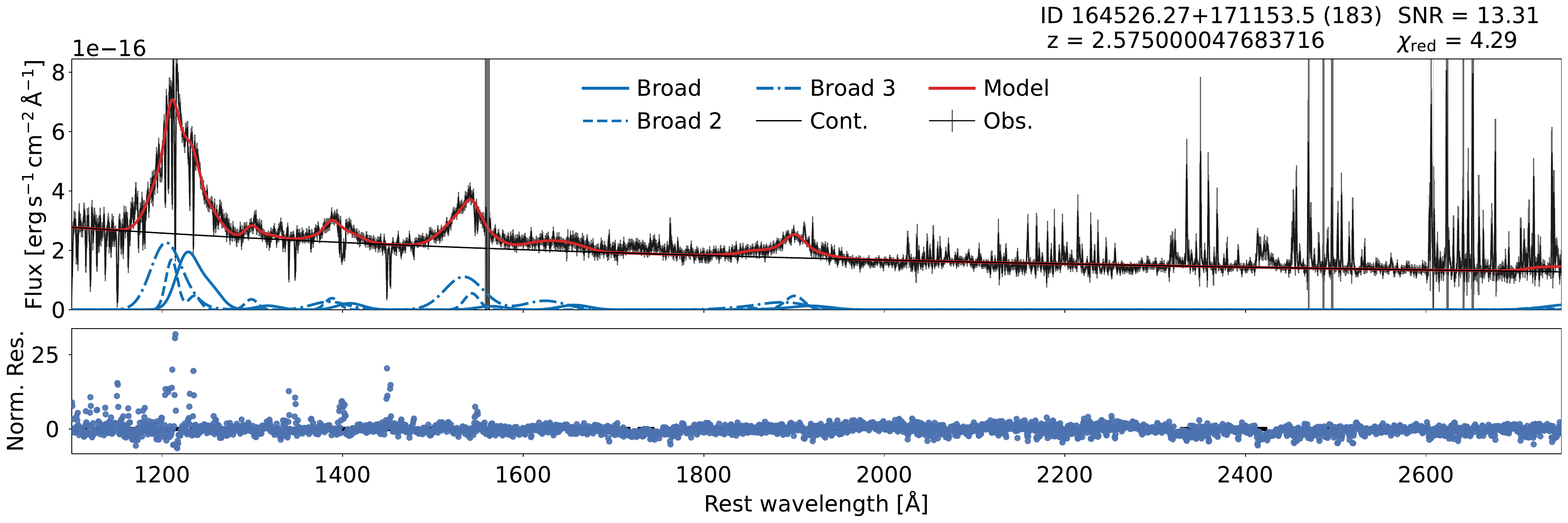}
    \caption{Example of fited spectra for the Wu\&Shen sample for the object with median $\chi^2_{\rm red}$}
    \label{fig:spectra_wushen_median_chi2}
\end{figure}

\begin{figure}[!htbp]
    \centering
    \includegraphics[width=1.0\linewidth]{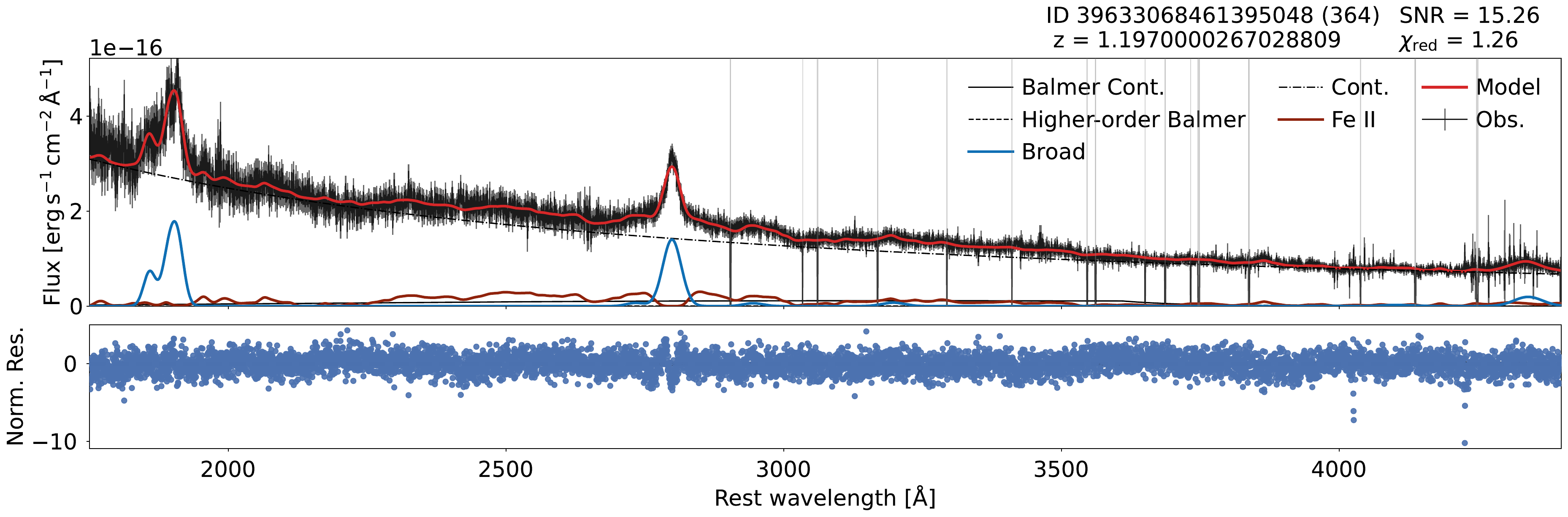}
    \caption{Example of fited spectra for the Pan+25 sample for the object with median S/N}
    \label{fig:spectra_pan_median_snr}
\end{figure}

\begin{figure}[!htbp]
    \centering
    \includegraphics[width=1.0\linewidth]{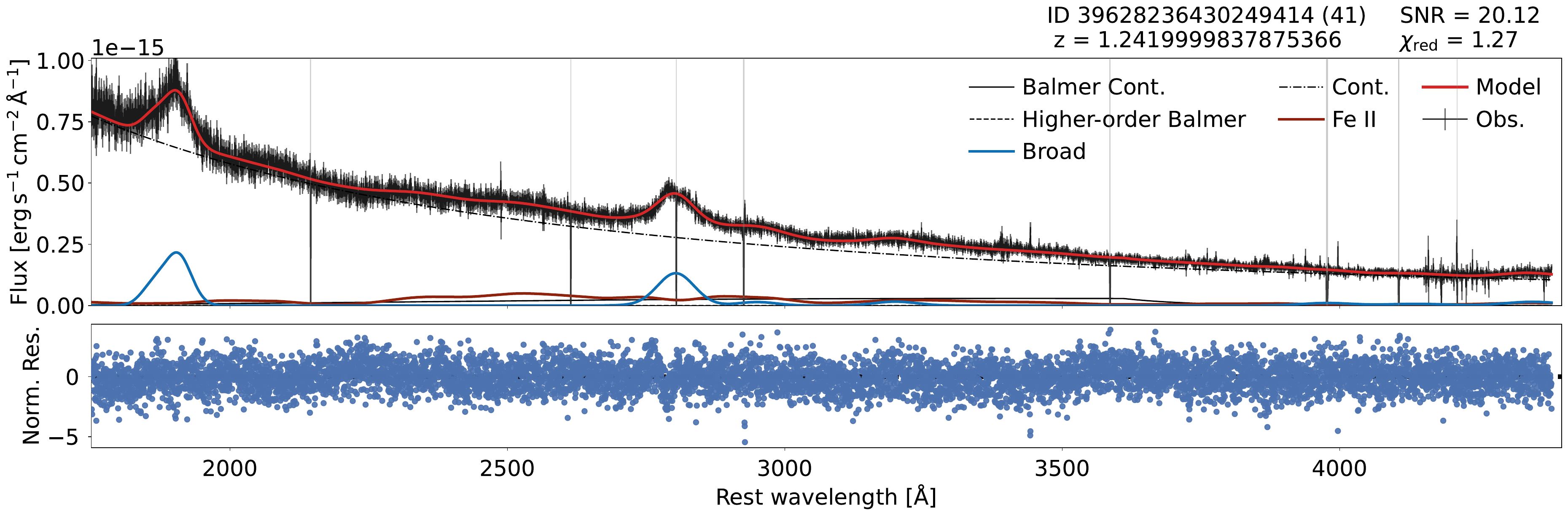}
    \caption{Example of fited spectra for the Pan+25 sample for the object with median $\chi^2_{\rm red}$}
    \label{fig:spectra_pan_median_chi2}
\end{figure}

\begin{figure}[!htbp]
    \centering
    \includegraphics[width=1.0\linewidth]{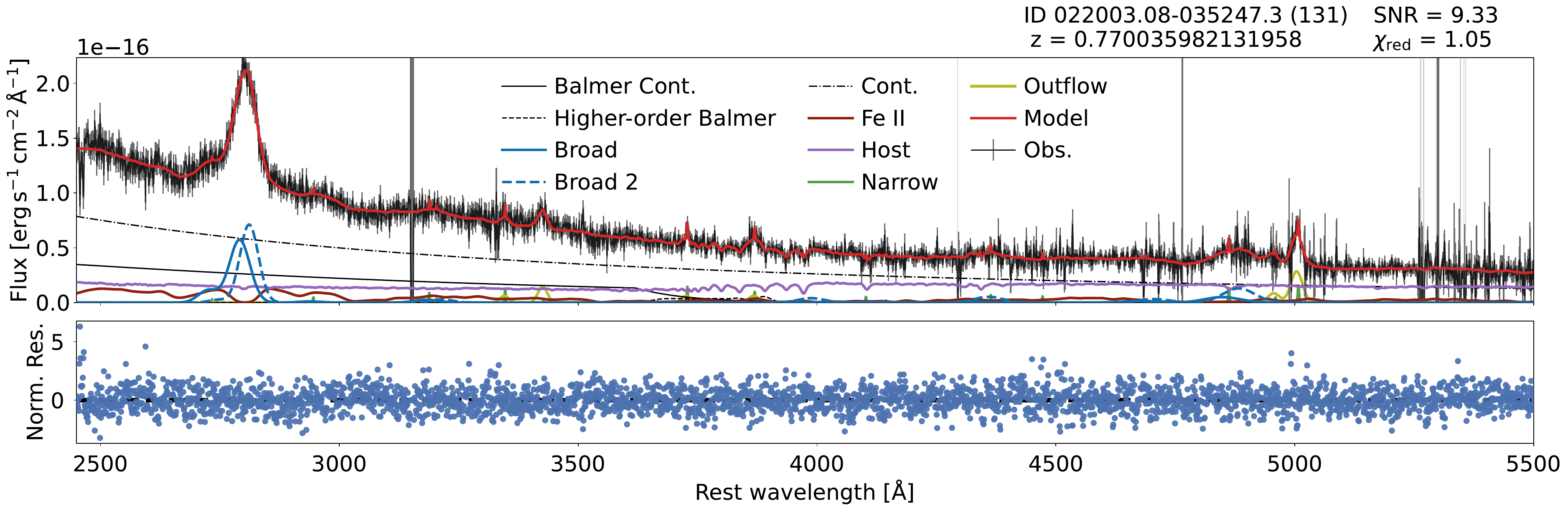}
    \caption{Example of fited spectra for the Sánchez-Sáez+18 sample for the object with median S/N}
    \label{fig:spectra_SS_median_snr}
\end{figure}

\begin{figure}[!htbp]
    \centering
    \includegraphics[width=1.0\linewidth]{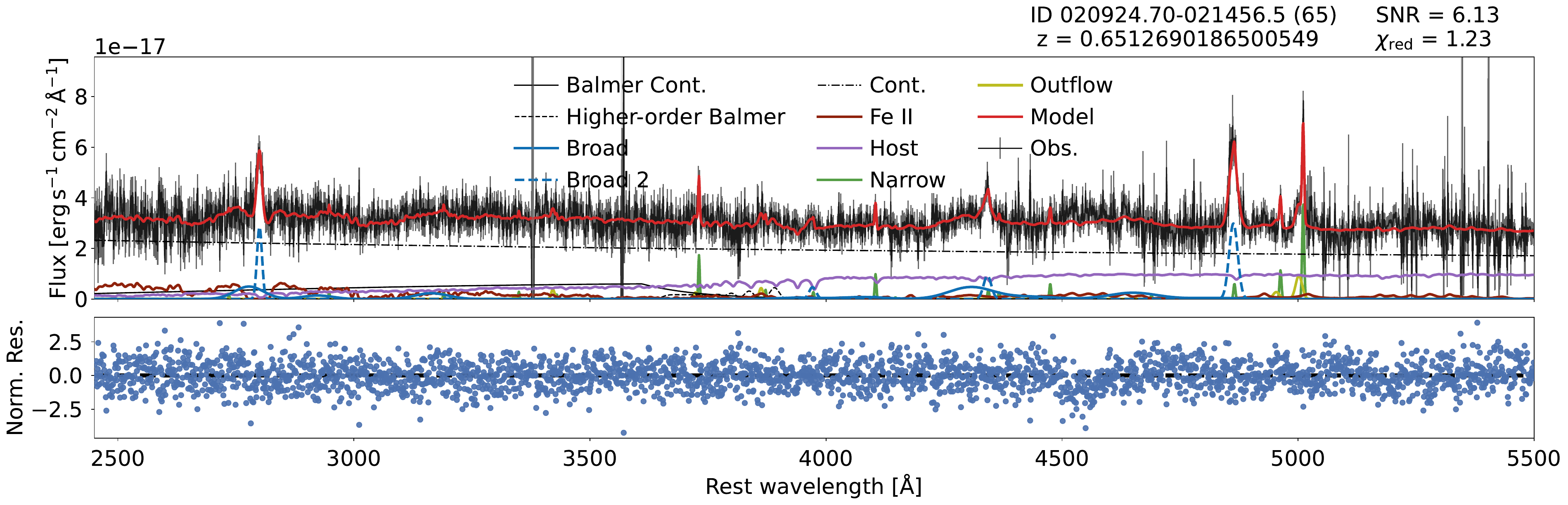}
    \caption{Example of fited spectra for the Sánchez-Sáez+18 sample for the object with median $\chi^2_{\rm red}$}
    \label{fig:spectra_SS_median_chi2}
\end{figure}

\begin{figure}[!htbp]
    \centering
    \includegraphics[width=1.0\linewidth]{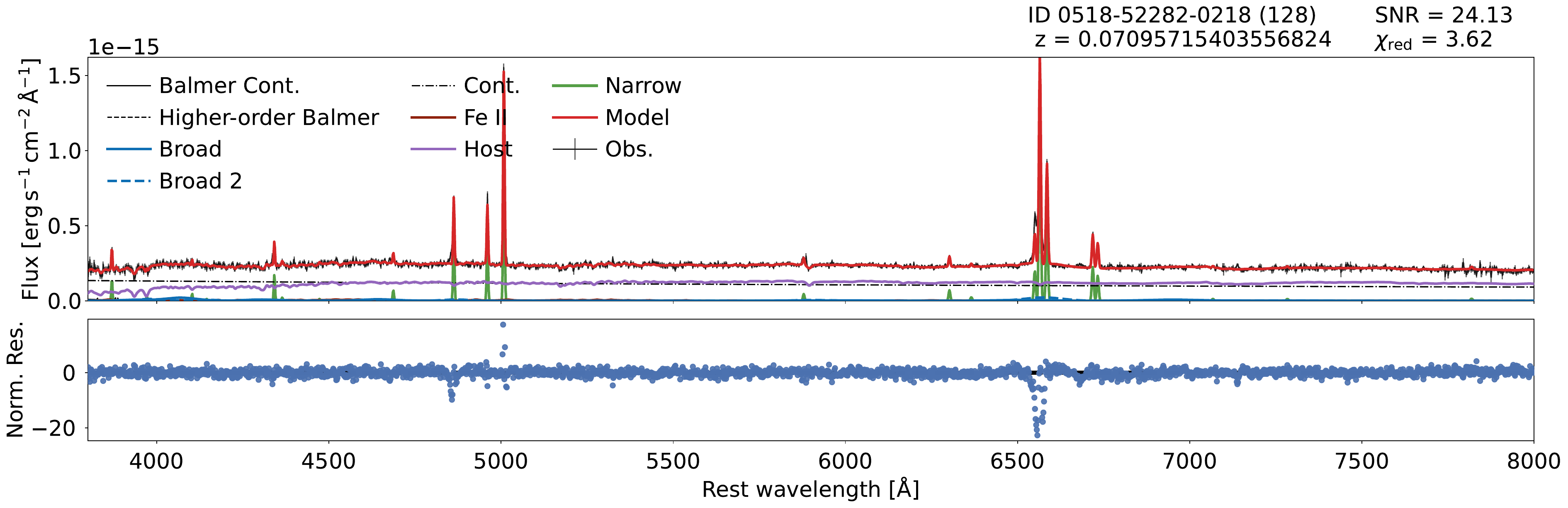}
    \caption{Example of fited spectra for the Bernal+25 sample for the object with median S/N}
    \label{fig:spectra_bernal_median_snr}
\end{figure}

\begin{figure}[!htbp]
    \centering
    \includegraphics[width=1.0\linewidth]{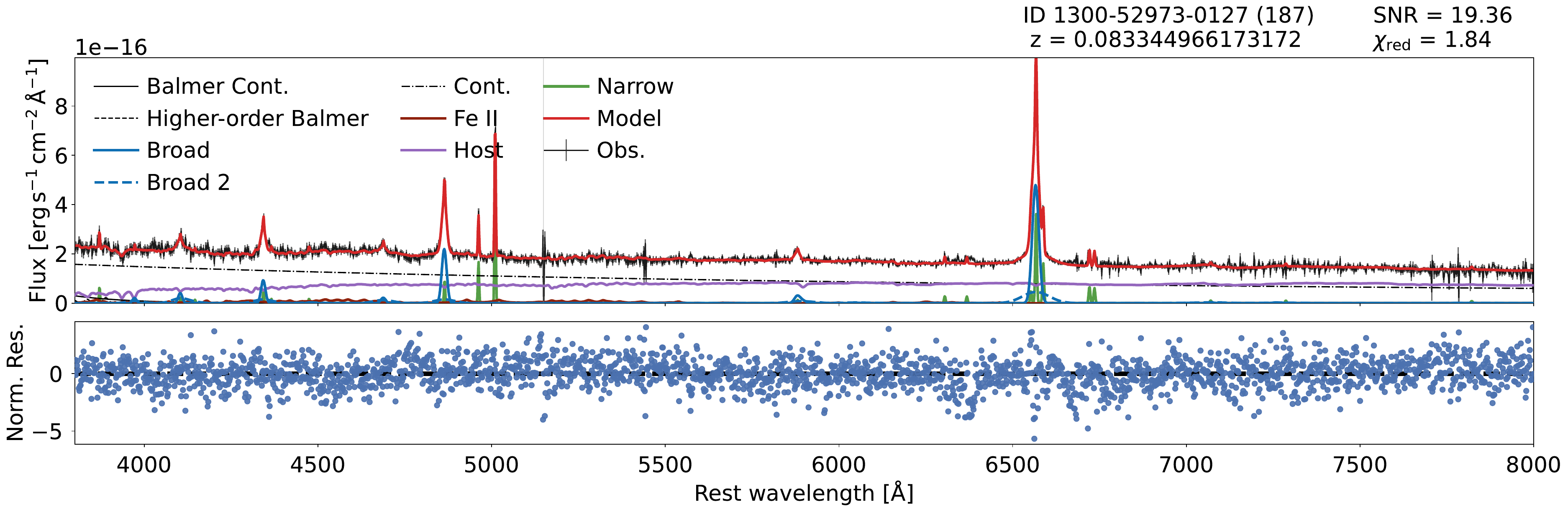}
    \caption{Example of fited spectra for the Bernal+25 sample for the object with median $\chi^2_{\rm red}$}
    \label{fig:spectra_bernal_median_chi2}
\end{figure}
\onecolumn
\section{Emission lines in the models}
\begin{table*}[!htbp]
\centering
\caption{Emission lines used by each the sample. Counts: Wu\&Shen22 (broad=3); Pan+25 (broad=1); Sánchez-Sáez+18 (broad=2, narrow=1, outflow=1); Bernal+25 (broad=2, narrow=1).}
\label{tab:emissionline}
\scriptsize
\begin{tabular}{lllll}
Region & Wu\&Shen22 & Pan+25 & Sánchez-Sáez+18 & Bernal+25 \\ [2pt]
\hline
broad & \begin{tabular}[t]{@{}l@{}} \ Ly$\alpha$~$\lambda\,1215.670$ \\\ NV~$\lambda\,1240.812$ \\\ OI+SiII~$\lambda\,1304.350$ \\\ SiIV~$\lambda\,1393.755$ \\\ OIV]~$\lambda\,1402.770$ \\\ CIV~$\lambda\,1549.479$ \\\ HeIIa~$\lambda\,1640.420$ \\\ AlIII~$\lambda\,1858.753$ \\\ SiIII]~$\lambda\,1892.000$ \\\ CIII]~$\lambda\,1908.730$ \\\ HeIIb~$\lambda\,2733.290$ \end{tabular} & \begin{tabular}[t]{@{}l@{}} \ AlIII~$\lambda\,1858.753$ \\\ SiIII]~$\lambda\,1892.000$ \\\ CIII]~$\lambda\,1908.730$ \\\ HeIIb~$\lambda\,2733.290$ \\\ MgII~$\lambda\,2799.000$ \\\ HeIa~$\lambda\,2945.106$ \\\ HeIb~$\lambda\,3187.745$ \\\ HeIIc~$\lambda\,3203.100$ \\\ H\ensuremath{\epsilon}~$\lambda\,3970.072$ \\\ H\ensuremath{\delta}~$\lambda\,4101.742$ \\\ HeIc~$\lambda\,4143.761$ \\\ Hg~$\lambda\,4340.471$ \end{tabular} & \begin{tabular}[t]{@{}l@{}} \ HeIIb~$\lambda\,2733.290$ \\\ MgII~$\lambda\,2799.000$ \\\ HeIa~$\lambda\,2945.106$ \\\ HeIb~$\lambda\,3187.745$ \\\ HeIIc~$\lambda\,3203.100$ \\\ H\ensuremath{\epsilon}~$\lambda\,3970.072$ \\\ H\ensuremath{\delta}~$\lambda\,4101.742$ \\\ HeIc~$\lambda\,4143.761$ \\\ Hg~$\lambda\,4340.471$ \\\ HeId~$\lambda\,4471.479$ \\\ HeIId~$\lambda\,4685.710$ \\\ H$\beta$~$\lambda\,4861.333$ \end{tabular} & \begin{tabular}[t]{@{}l@{}} \ H\ensuremath{\epsilon}~$\lambda\,3970.072$ \\\ H\ensuremath{\delta}~$\lambda\,4101.742$ \\\ HeIc~$\lambda\,4143.761$ \\\ Hg~$\lambda\,4340.471$ \\\ HeId~$\lambda\,4471.479$ \\\ HeIId~$\lambda\,4685.710$ \\\ H$\beta$~$\lambda\,4861.333$ \\\ HeIe~$\lambda\,5877.254$ \\\ NaIa~$\lambda\,5890.000$ \\\ NaI1a~$\lambda\,5896.000$ \\\ OIa~$\lambda\,6046.440$ \\\ H$\alpha$~$\lambda\,6562.820$ \\\ OIb~$\lambda\,7002.230$ \\\ HeIf~$\lambda\,7065.196$ \\\ OIc~$\lambda\,7254.448$ \\\ HeIg~$\lambda\,7281.349$ \\\ OId~$\lambda\,7774.170$ \\\ HeIh~$\lambda\,7816.136$ \end{tabular} \\ [2pt]
\hline
narrow & -- & -- & \begin{tabular}[t]{@{}l@{}} \ HeIIb~$\lambda\,2733.290$ \\\ HeIa~$\lambda\,2945.106$ \\\ HeIb~$\lambda\,3187.745$ \\\ HeIIc~$\lambda\,3203.100$ \\\ [NeV]a~$\lambda\,3346.790$ \\\ [NeV]b~$\lambda\,3426.850$ \\\ [OII]~$\lambda\,3728.000$ \\\ [NeIII]~$\lambda\,3868.760$ \\\ H\ensuremath{\epsilon}~$\lambda\,3970.072$ \\\ H\ensuremath{\delta}~$\lambda\,4101.742$ \\\ HeIc~$\lambda\,4143.761$ \\\ Hg~$\lambda\,4340.471$ \\\ OIIIa~$\lambda\,4363.214$ \\\ HeId~$\lambda\,4471.479$ \\\ HeIId~$\lambda\,4685.710$ \\\ H$\beta$~$\lambda\,4861.333$ \\\ OIIIb~$\lambda\,4958.896$ \\\ OIIIc~$\lambda\,5006.803$ \end{tabular} & \begin{tabular}[t]{@{}l@{}} \ [NeIII]~$\lambda\,3868.760$ \\\ H\ensuremath{\epsilon}~$\lambda\,3970.072$ \\\ H\ensuremath{\delta}~$\lambda\,4101.742$ \\\ HeIc~$\lambda\,4143.761$ \\\ Hg~$\lambda\,4340.471$ \\\ OIIIa~$\lambda\,4363.214$ \\\ HeId~$\lambda\,4471.479$ \\\ HeIId~$\lambda\,4685.710$ \\\ H$\beta$~$\lambda\,4861.333$ \\\ OIIIb~$\lambda\,4958.896$ \\\ OIIIc~$\lambda\,5006.803$ \\\ HeIe~$\lambda\,5877.254$ \\\ OIa~$\lambda\,6300.304$ \\\ OIb~$\lambda\,6363.776$ \\\ NIIa~$\lambda\,6548.050$ \\\ H$\alpha$~$\lambda\,6562.820$ \\\ NIIb~$\lambda\,6583.460$ \\\ SIIa~$\lambda\,6716.440$ \\\ SIIb~$\lambda\,6730.810$ \\\ HeIf~$\lambda\,7065.196$ \\\ HeIg~$\lambda\,7281.349$ \\\ HeIh~$\lambda\,7816.136$ \end{tabular} \\ [2pt]
\hline
outflow & -- & -- & \begin{tabular}[t]{@{}l@{}} \ [NeV]a~$\lambda\,3346.790$ \\\ [NeV]b~$\lambda\,3426.850$ \\\ [OII]~$\lambda\,3728.000$ \\\ [NeIII]~$\lambda\,3868.760$ \\\ OIIIb~$\lambda\,4958.896$ \\\ OIIIc~$\lambda\,5006.803$ \end{tabular} & -- \\ [2pt]
\end{tabular}
\end{table*}

\end{appendix}

\end{document}